\documentclass[reprint,amsmath,amssymb,aps]{revtex4-2}

\usepackage{floatrow}
\usepackage{graphicx}% Include figure files
\usepackage{dcolumn}% Align table columns on decimal point
\usepackage{bm}% bold math
\usepackage{xcolor}
\usepackage{hyperref}% add hypertext capabilities
\usepackage{tcolorbox}
\usepackage[top=0.5in,bottom=0.5in,left=0.5in,right=0.5in]{geometry}

\renewcommand{\thesubsection}{\arabic{section}.\arabic{subsection}}
\makeatletter
\renewcommand{\p@subsection}{}
\renewcommand{\p@subsubsection}{}
\makeatother

\def\f{\frac}
\newcommand\fp[2]{\left(\frac{{#1}}{{#2}}\right)}
\renewcommand\Im[1]{\tn{Im}\left[#1\right]}
\renewcommand\Re[1]{\tn{Re}\left[#1\right]}
\def\ds{\displaystyle} % used for display math in inline mode
\def\l{\left}
\def\r{\right}
\def\d{\mathrm{d}}
\def\p{\partial}
\def\<{\langle}
\def\>{\rangle}
\def\al{\alpha}
\def\ch{\chi}
\def\dl{\delta}
\def\om{\omega}
\def\rh{\rho}
\def\lm{\lambda}
\def\ph{\phi}
\def\th{\theta}
\def\ga{\gamma}
\def\Om{\Omega}
\def\bt{\beta}
\def\et{\eta}
\def\div{\nabla}
\def\et{\eta}
\def\LL{\mathcal{L}}
\def\diss{\tn{Diss}}
\def\costfn{\mathcal{C}}
\def\dissfn{\mathcal{D}}
\def\tn{\textnormal}
\def\sg{\sigma}
\def\sgI{\sigma_I}
\def\sgO{\sigma_O}
\def\CC{\mathcal{C}}
\def\OO{\mathcal{O}}

\def\LL{\mathcal{L}}
\def\FF{\mathcal{F}}

\def\erfc{\tn{erfc}}
\def\bb{\boldsymbol}
\renewcommand{\t}[1]{\tilde{#1}}
\newcommand{\kbt}{k_B\hspace*{-1.5pt}T}
\def\el{\tn{El}}
\def\dff{\tn{D2}}
\def\dfff{\tn{D3}}
\def\ac{\tn{Ac}}
\begin{document}

\title{Physical constraints in intracellular signaling: the cost of sending a bit}

\author{Samuel J. Bryant}
 \email{samuel.bryant@yale.edu}
 \affiliation{Department of Physics, Yale University.}

\author{Benjamin B. Machta}%
 \email{benjamin.machta@yale.edu}
\affiliation{Department of Physics, Yale University}%
\affiliation{Systems Biology Institute, Yale University}%

\date{\today}

\begin{abstract}
One of the primary computational requirements of a cellular system is the ability to transfer information between spatially separated components.
To accomplish this, biology uses diverse physical channels including production or release of second-messengers molecules and electrical depolarization of the plasma membrane.
To send reliable information, these processes must dissipate energy to compete with thermal noise, in some cases consuming a substantial fraction of the cellular energy budget.
Here we bound the energetic efficiency of several physical strategies for communication, using tools from information theory and the fluctuation dissipation relations to quantify communication through a channel corrupted by thermal noise.
We find a minimum energetic cost, in $k_B$T/bit for sending information as a function of the size of the sender and receiver, their spatial separation, and the communication latency.
From these calculations construct a phase diagram indicating where each strategy is most efficient.
In addition, these calculations provide an estimate for the energy costs associated with information processing arising from the physical constraints of the cellular environment.
\end{abstract}

\maketitle

\section{Background}

All life must couple entropically unfavorable changes to the dissipation of energy.
Some of this energy is consumed to perform mechanical tasks for which thermodynamics provides a conceptual framework for understanding energy consumption.
However, a large portion of this energy consumption goes towards information processing, broadly defined: controlling molecular machines, synchronizing biochemical networks, and coherently responding to stimuli.
A large body of work has aimed to theoretically bound the energy needed for many types of information processing, including the cost of precisely reading DNA \cite{hopfield_kinetic_1974},
performing abstract computation \cite{landauer_irreversibility_1961,bennett_thermodynamics_1982,wolpert_stochastic_2019,chen_deterministic_2013,kolchinsky_entropy_2021},
measuring and sensing the environment \cite{mehta_energetic_2012,govern_energy_2014,wang_price_2020,ouldridge_thermodynamics_2017,sartori_thermodynamic_2014,barato_information-theoretic_2013},
breaking time-reversal symmetry \cite{feng_length_2008,parrondo_entropy_2009,barato_thermodynamic_2015,brown_effective_2016,rao_nonequilibrium_2016}
keeping accurate time \cite{cao_free-energy_2015,barato_cost_2016,zhang_energy_2019}, self-replication \cite{england_statistical_2013} and controlling a small thermodynamic system \cite{sivak_thermodynamic_2012,Machta15,bryant_energy_2020}.
The abstract nature of these bounds makes them broadly applicable, but often at the cost of divorcing them from the details of their physical implementations. For many cellular examples, the bounds appear dramatically far from saturated \cite{rodenfels_heat_2019,Levy21,Attwell01,Laughlin98}.

However, biological systems are subject to constraints often not captured in these theoretical abstractions.
In particular, information processing networks are by nature distributed in space, and therefore it is necessary to send information over physical distances under specific time constraints.
For example, chemoreceptors in bacteria measure environmental information which must then travel $\sim 1$ $\mu$m from receptor clusters to cellular motors within a fraction of a second to be behaviorally useful.
In neurons, information arriving at synapses in distant dendrites must travel across the cell body, perhaps a millimeter in length, in timescales of milliseconds.

The schemes that have evolved to move information across space quickly are varied, not just in their molecular details, but in their underlying physics.
In neurons, signals are transmitted electrically via the opening of ion channels which depolarize the membrane, causing distant changes in electrical potential.
All cells signal through the creation, release or modification of second messenger molecules which diffuse to distant targets.
At the organism level, pressure waves transmit information over longer distances in the form of sound.

While moving information is not a process that has a fundamental energetic cost, the practical costs can be substantial - a large fraction~\cite{smith_cerebral_2002,raichle_appraising_2002} of the energy humans consume is spent by neurons to generate voltage gradients, primarily used for sending signals over long distances~\cite{Levy21}.

\section{Overview of Work}

In this manuscript we estimate bounds for this energetic cost of sending information.  We examine several physical communication strategies used by biology): 1) electrical signaling via the depolarization of membranes through ion channels, 2) diffusive signaling in 2D and 3D, and 3) acoustic signaling.
The resulting energetic costs, in $k_BT$ per bit, depend on four key parameters: the distance the signal is sent $r$, the signal frequency $\om$, and the size of the sender/receiver $\sgI$/$\sgO$.
The bounds we find do not represent fundamental costs associated with information processing, but instead represent costs associated with the constraints that biology faces, subject to life's practical existence in a watery buffer.
As such our bounds contain not just pure numbers and the thermal energy $k_B T$, but also physical diffusion constants, the plasma membrane's capacitance, and the cytoplasm's electrical conductivity and viscosity.

We find that each physical  strategy is characterized by a functional scaling form that describes the cost  of sending information over a distance, with particular dependence on the size of the sensors.
Each strategy also has a characteristic lengthscale $\ell$, which depends on its associated physical constants and the frequency of the signal, setting an upper limit on how far it can be used efficiently.
When the transmission distance $r$ exceeds this characteristic lengthscale $\ell$, the signal becomes exponentially costly to send.
From these characteristic lengthscales, we can determine when certain signaling schemes are optimal and when they are forbidden (fig.~\ref{fig:phase-diagram}b).

Like previous work, we have used thermodynamic tools to place energetic bounds on information processing at the cellular level.
However, our approach is novel in considering the practical constraints imposed by the physical environments available to biology.
Our results show that there are large energetic costs associated with biological function that cannot be obtained from purely abstract thermodynamic and information theoretic considerations.
We believe that this approach will help bridge the gap between the experimentally observed energetic usage and the weak energetic bounds found in the theoretical community.

\begin{figure*}[ht]
  \centering
  \parbox{\textwidth}{
    \parbox{.365\textwidth}{%
      \includegraphics[width=0.365\textwidth]{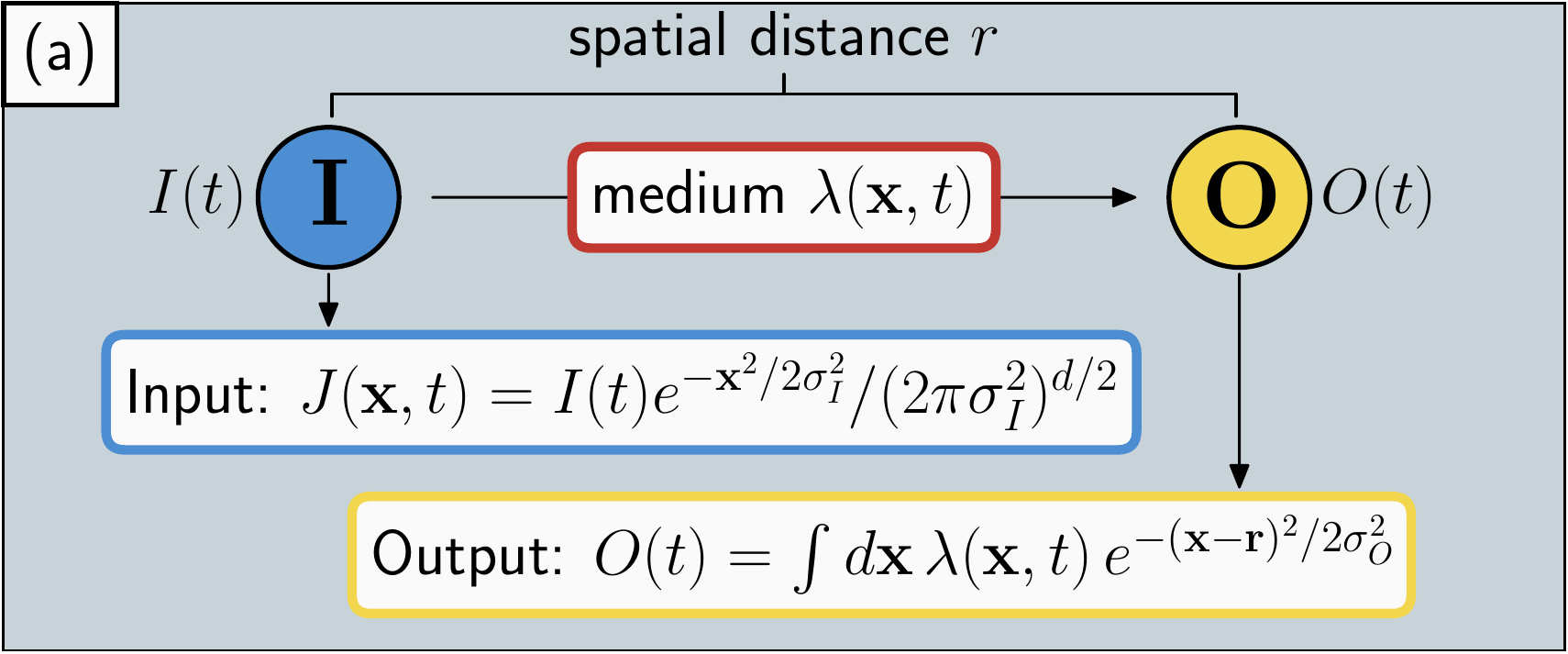}
      \includegraphics[width=0.365\textwidth]{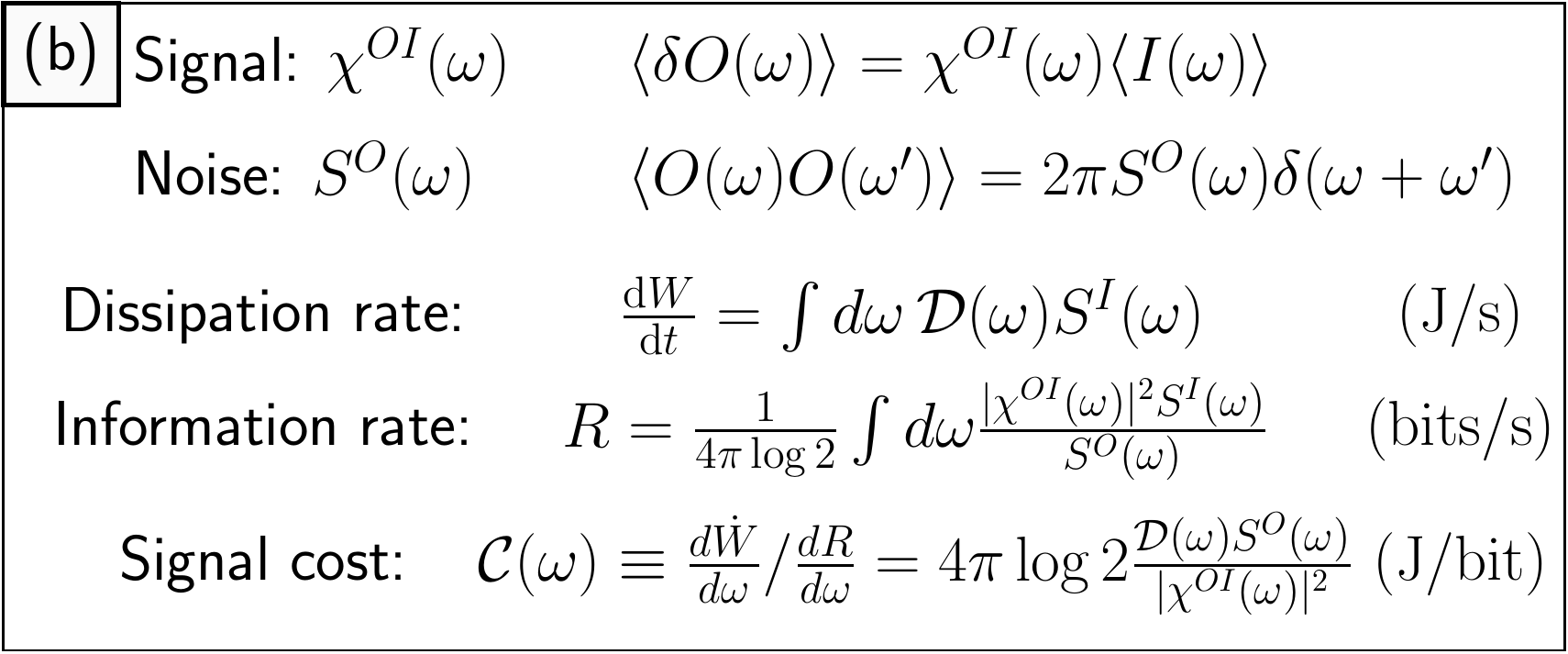}
    }
    \parbox{.3\textwidth}{%
      \includegraphics[width=0.3\textwidth]{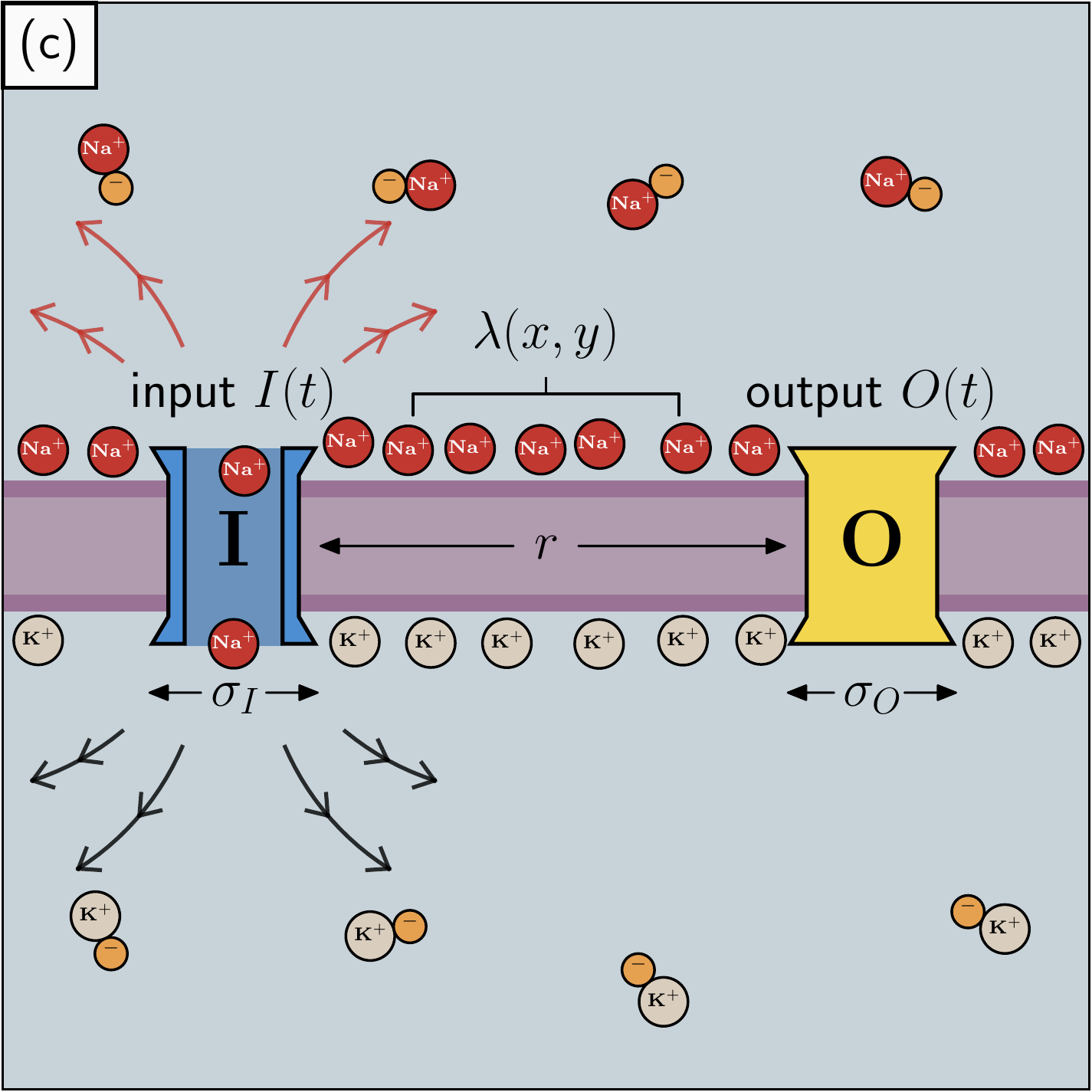}
    }
    \parbox{.3\textwidth}{%
      \includegraphics[width=0.3\textwidth]{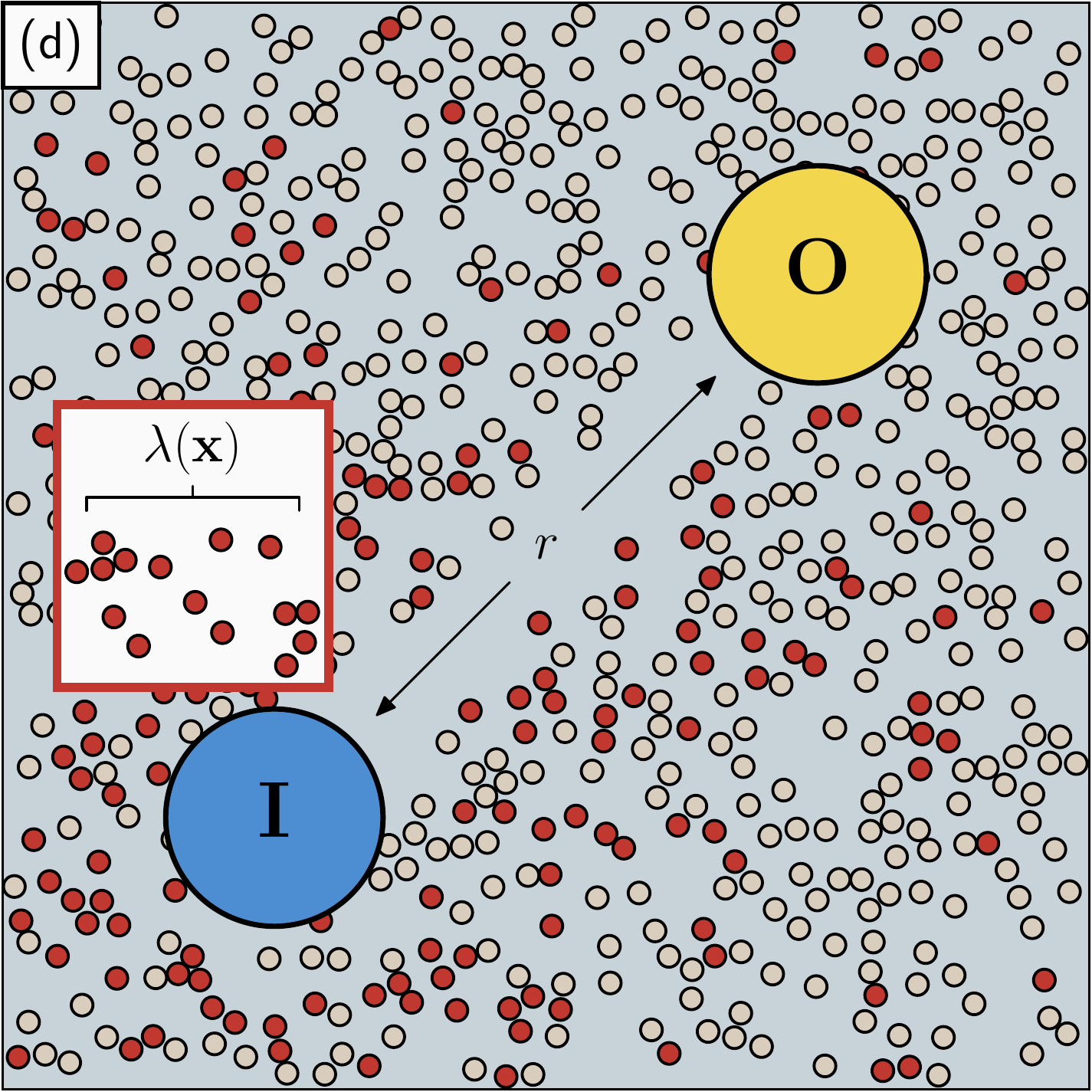}
    }
  }
  \caption{
    \label{fig:fig1}
    An overview of schemes for sending information across spatial distances.
    (a) An abstract transmission scheme which sends information from a sender, $I$, to a receiver, $O$. The signal is propagated via a coupling to a density field $\lambda({\bf x},t)$. The sender generates a perturbation $I(t)$ which acts on $\lm$ through the current density $J$. The receiver observes the signal by measuring $O(t)$, the localized change in the density field. (b) The signal transmission strength, the noise, and the dissipation rate required to produce the signal in terms of the fundamental fields. The signal transmission coefficient $\ch^{OI}$ is the linear response kernel of $O$ to $I$. The noise $S^O(\om)$ is defined as the equilibrium fluctuations in $O(t)$, computed using the fluctuation dissipation theorem~\cite{tong_lectures_2012}. The dissipation rate is system-specific, presented here in terms of the dissipation $\dissfn(\om)$ at each frequency.
    (c) A depiction of electrical signaling between two ion channels in a membrane. The input signal, $I(t)$, is the time-varying current flowing through the sender ion channel. The output, $O(t)$, is the excess charge accumulated at the receiver. The density field, $\lambda$, is the surface charge density on the membrane.
    (d) A depiction of a diffusive signaling system between two proteins embedded in a 2D membrane. The input signal, $I(t)$, is the time-varying rate of particle injection at the sender.
  }
\end{figure*}

\section{Energetic cost of sending information}
\label{sec:cost_per_bit}

Here we introduce the energetic-information framework (summarized in fig.~\ref{fig:fig1}a-b) we use to compute the cost of sending information across physical space.
We suppose that there is a sender, $I$, and a receiver, $O$, embedded in a background medium $\lambda({\bf x}, t)$ subject to thermal fluctuations.
The sender transmits a time-varying signal $I(t)$ by locally perturbing the medium $\lm$.
The receiver then measures the signal by observing $O(t)$, the local deviation of the medium $\lm$ from equilibrium.
From this measurement, the receiver is able to extract information about the state of the sender.
We refer to $I(t)$ and $O(t)$ as the input and output.
The locality of the input and output behavior is quantified by a Gaussian density profile, as defined in fig. \ref{fig:fig1}a.
$\lm$ is a density field, whose nature and dynamics depend on the specific communication medium (see table ~\ref{tab:results}).

We define the rate of information transfer between the sender and receiver as the time-series mutual information rate between $I(t)$ and $O(t)$~\cite{tostevin_mutual_2009}, measured in bits per second.
Following \cite{tostevin_mutual_2009}, we assume that we are in the weak signal regime and further that noise is dominated by thermal fluctuations in $\lm$.
With this assumption, the rate that information is sent from the input to the output (in bits/sec) is given by:
\begin{equation}
  \label{eq:info_rate}
  R(I,O)
  \approx
  \frac{1}{4\pi\log 2}
  \int\d\om\frac{|\ch^{OI}(\om)|^2 S^{I}(\om)}{S^{O}(\om)}
\end{equation}
where $|\ch^{OI}|^2$ and $S^{O}(\om)$ correspond to the transmission gain and noise, as defined in fig. \ref{fig:fig1}b, and where $S^I(\om)$ is the power spectrum of the signal process, chosen by the sender, and thus external to the network itself.

To compute the energetic cost of signaling, we need to compare the information rate to the rate of work required to produce the input signal. We characterize this dissipation rate $\dot{W}$ by a dissipation kernel function $\dissfn(\om)$, describing the rate of energy dissipation at each signal frequency (fig. \ref{fig:fig1}b). If we consider signals localized in frequency space, we can describe the overall cost per bit to send a signal at frequency $\omega$ as the ratio of the dissipation rate integrand to the information rate integrand (see methods sec.~\ref{sec:methods:cost-per-bit}):
\begin{equation}
  \label{eq:cost_per_bit}
  \costfn(\om)
  \equiv
  \frac{\tn{Cost}}{\tn{bit}}
  \equiv
  \frac{\d\dot{W}}{\d\om}\left/\frac{\d R}{\d\om}\r.
  =
  4\pi\log 2\frac{\dissfn(\om)S^{O}(\om)}{|\ch^{OI}(\om)|^2}
\end{equation}

To compute this quantity for a given model, we need the equilibrium noise spectrum $S^{O}(\om)$, the dissipation kernel $\dissfn(\om)$, and the transmission coefficient $\ch^{OI}(\om)$.
In section \ref{sec:electrical}, we sketch this analysis explicitly for the case of electrical signaling in membranes to illustrate the techniques.
The analysis for the other systems (diffusive and acoustic signaling) follows similar procedures and details are provided in the methods section with key intermediates summarized in table~\ref{tab:results}.
A complete derivation of this result for all systems, including asymptotic behavior, can be found in appendix \ref{app:sec:electrical}-\ref{app:sec:acoustic}.

\section{Application to Electrical Signaling Between Ion Channels}
\label{sec:electrical}

We suppose there is a signaling process between two membrane-bound ion channels embedded in a 2D membrane, as depicted in fig.~\ref{fig:fig1}c.
The membrane is an infinite 2D capacitor with capacitance $c$ (F/m$^2$) and surface charge density $\lm(x,y)$.
It's embedded within a bulk conductor extending into the $z$ direction, with conductance $\al$ ($\Om$/m).
Free charges can accumulate along the membrane but may not exist in the bulk.
Following fig \ref{fig:fig1}a, the input of the system, $I(t)$, is the time-varying flow of current through the sender ion channel located at the origin, spread out over a  2D Gaussian disk of radius $\sgI$.
The output of the system, $O(t)$, is the excess charge measured over a Gaussian disk of radius $\sgO$ at the receiver ion channel located at $\bb r$.
Here we assume an infinite flat membrane, though the geometry of specific systems is likely important, for example in roughly cylindrical axons and dendrites.

\subsection{Linearized dynamics}

To compute the quantities needed for eq. (\ref{eq:cost_per_bit}), we first need a minimal model for the dynamics of $\lm$.
The behavior of the bulk is described by the potential $V(x,y,z)$ obeying Laplace's equation $\nabla^2 V=0$ for $z>0$.
At the membrane, the voltage is given by the local capacitance equation: $V(x,y,z=0)=\lm(x,y)/c - h(x,y)$ where $h$ is an artificial external field useful in calculating the spectrum of thermal charge fluctuations.
Bulk current flows according to  $-\al\nabla V$, which is divergenceless everywhere except at the membrane.
Thus, the rate of change of charge at the membrane is given by the sum of the injected current, $J(x,y)$, and the rate that charge accumulates from bulk currents: $\partial_t \lm(x,y) = J(x,y)+\alpha \partial_z V(x,y,z)|_{z=0}$.
These equations are linear, and in $xy$-Fourier space they close in terms of $\lm$ and applied fields $h$ and $J$ yielding:
\begin{equation}
  \label{eq:electrostatic_dynamics}
  \p_t\lm(\bb k,t) = -\al k\l[\frac{\lm(\bb k,t)}{c} - h(\bb k,t)\r]
  + J(\bb k,t)
\end{equation}
Here $\bb k=(k_x,k_y)$ is the $xy$-momentum vector and all given quantities are implicit Fourier transforms.
This result will be used to derive the transmission coefficient, $\ch^{OI}$ (the linear response function of $O$ to $I$~\cite{zwanzig_nonequilibrium_2001,tong_lectures_2012}), the dissipation kernel, $\dissfn(\om)$, and, in conjunction with the fluctuation dissipation theorem, the noise in the output signal, $S^{O}(\om)$~\cite{zwanzig_nonequilibrium_2001,tong_lectures_2012}.

\subsection{Calculating Transmission Strength}

The transmission strength is characterized by the linear response function $\ch^{OI}$ which indicates how the mean output $\<O(t)\>$ responds to the input $I(t)$.
As an intermediary step, we first compute the response of the charge density to the input signal, $\ch^{\lm J}$ in frequency space by reading off the frequency-space Fourier transform of eq. (\ref{eq:electrostatic_dynamics}): $\chi^{\lm J}(k,\omega)=c/\al(|k|+i\om c/\al)$.
By then integrating $\ch^{\lm J}$ over the sensor area, and making the assumption that $r \gg\sgO$, $\sgI$, we get the transmission coefficient $\ch^{OI}$:
\begin{align}
  \label{eq:electrostatic_signal}
  |\ch^{OI}(\om)|^2 = \frac{c^2\sgO^4}{\al^2 r^2}U_S\l(\frac{r}{\ell(\om)}\r)
  &&
  \ell(\om) = \frac{\al}{\om c}
\end{align}
where $U_S$ is a universal function of its argument that goes to $1$ when $r/\ell(\om) \ll 1$ and then exponentially decays (see appendix \ref{app:sec:electrical} for a more complete derivation including its explicit form and asymptotic behavior).
Very importantly, we have expressed this universal function in terms of the lengthscale $\ell(\om)=\al/\om c$ which sets an upper limit on the viability of transmission.
When the transmission distance $r$ exceeds $\ell(\om)$, $U_S$ falls off exponentially.
The origin of this lengthscale is related to the RC timescale found in basic RC circuits.
The 3D bulk resistance and 2D membrane capacitance together define an RC (inverse) {\it velocity}, or equivalently, a lengthscale at a given frequency.

\subsection{Calculating Dissipation}

In the linear response regime the instantaneous dissipation associated with powering a trans-membrane current is given by a spatial integral of the injected current density multiplied by the voltage across the membrane.  In frequency space this can be calculated from the response function, yielding:
\begin{equation}
  \label{eq:electrostatic_diss}
  \dissfn(\om) = \frac{1}{8\pi^{3/2}\al\sgI}U_D\l(\frac{\sgI}{\ell(\om)}\r)
\end{equation}
where $U_D$ is another universal function which goes to $1$ when $\sgI \ll \ell(\om)$.

\subsection{Calculating Noise}

We define the noise $S^{O}(\om)$ to be the power spectrum of the equilibrium fluctuations in the output $O(t)$ in the absence of an input signal.
As with $\ch^{OI}$, we first compute $\ch^{\lm h}$, the susceptibility of the charge density field to the external field $h$, by reading off eq. (\ref{eq:electrostatic_dynamics}): $\ch^{\lm h}(\bb k,\om) = \al k/(\al k/c + i\om)$.
The fluctuation dissipation theorem~\cite{zwanzig_nonequilibrium_2001,tong_lectures_2012} then tells us that the equilibrium fluctuations of $\lm$ are related to the imaginary part of $\ch^{\lm h}$: $S^\lm(\om)=-(2\kbt/\om)\tn{Im}(\ch^{\lm h})$.
We then integrate $S^\lm$ over the sensor area to obtain $S^{O}(\om)$:
\begin{equation}
  \label{eq:electrostatic_noise}
  S^{O} = \frac{2\pi^{3/2}\kbt c^2\sgO^3}{\al}U_N\l(\frac{\sgO}{\ell(\om)}\r)
\end{equation}
where $U_N$ is another universal function which goes to $1$ when $\sgO \ll \ell(\om)$.

\subsection{Energetic Cost per Bit}

Plugging these results into the cost per bit (eq. (\ref{eq:cost_per_bit})) yields the energetic cost of sending a signal at frequency $\omega$ over a distance $r$ between two ion channels:
\begin{align}
  \label{eq:electrostatic_result}
  \costfn^{\el} = \pi\log 2\frac{r^2}{\sgI\sgO}\,U^\el\l(\frac{r}{\ell(\om)}\r)
  \qquad
  (\kbt/\tn{bit})
\end{align}
We call $\pi r^2/\sgI\sgO$ the {\it scaling} portion of the cost and $U^\el$ the lengthscale correction which can be ignored when the transmission distance, $r$, is smaller than the characteristic lengthscale, $\ell(\om)$
(the universal functions $U_D$, $U_N$ from eq. (\ref{eq:electrostatic_diss}), (\ref{eq:electrostatic_noise}) can be ignored when $r<\ell(\om)$ since $\sgI$, $\sgO \ll r$ in any realistic signaling process).
Thus, when the transmission distance is {\it below} the characteristic lengthscale ($ r \ll \ell(\om) $), the cost per bit is independent of all the system constants and transmission frequency; it depends only on the lengthscales $r$, $\sgI$, $\sgO$.
However, as illustrated in fig. \ref{fig:phase-diagram}c, when the distance {\it exceeds} the characteristic lengthscale ($r > \ell(\om)$), the universal function $U^\el$ blows up, and the cost per bit rises exponentially.
We thus say that the characteristic lengthscale $\ell(\om)$ draws an exclusion zone (fig. \ref{fig:phase-diagram}b), a feature found in all of the systems we studied.

\begin{table*}
  \begin{center}
  \begin{tabular}{ |c|c|c|c|c| }
    \hline
  System / Example
    &
    Coupling field $\lambda$ / Dynamics
    &
    Input $I(t)$ / Output $O(t)$
    &
    Energy cost $\mathcal{C}$
    &
    Lengthscale \\
  \hline
    \begin{tabular}{c}
      {\bf Electrical}
      \\
      ion channels
      \\ in neurons
    \end{tabular}
    &
    \begin{tabular}{c}
      surface charge density (C/m$^2$)\vspace*{4pt}
      \\
      $\p_t\lm(\bb k,t) = -\frac{\al k}{c}\lm(\bb k,t)$
    \end{tabular}
    &
    \begin{tabular}{c}
      $I(t) =$ injected current ($\tn{A}$)
      \vspace*{4pt}
      \\
      $O(t) =$ excess charge (C)
    \end{tabular}
    &
    $\ds\costfn^\el=\pi\log 2\frac{r^2}{\sgI\sgO}$
    &
    $\ds\ell(\om)=\frac{\al}{\om c}$\\
    \hline
  \begin{tabular}{c}
    {\bf Diffusion 2D}
    \\
    Pip2
  \end{tabular}
    &
    \begin{tabular}{c}
      messenger density (1/m$^2$)
      \vspace*{4pt}
      \\
      $\p_t\lm(\bb x,t) = D\nabla^2\lm(\bb x,t)$
    \end{tabular}
    &
    \begin{tabular}{c}
      $I(t) =$ activation rate (Hz)
      \vspace*{4pt}
      \\
      $O(t) =$ messenger count (1)
    \end{tabular}
    &
    $\ds\costfn^\dff \approx 4\log 2\frac{\log\l(\frac{\ell}{\sgI}\r)\log\l(\frac{\ell}{\sgO}\r)}{\log\l(\frac{\ell}{r}^2\r)}$
    &
    $\ds\ell(\om)=\sqrt{\frac{D}{\om}}$\\
    \hline
  \begin{tabular}{c}
    {\bf Diffusion 3D}
    \\
    CheY in E coli
  \end{tabular}
    &
    \begin{tabular}{c}
      messenger density (1/m$^3$)\vspace*{4pt}
      \\
      $\p_t\lm(\bb x,t) = D\nabla^2\lm(\bb x,t)$
    \end{tabular}
    &
    \begin{tabular}{c}
      $I(t) =$ activation rate (Hz)
      \vspace*{4pt}
      \\
      $O(t) =$ messenger count (1)
    \end{tabular}
    &
    $\ds \costfn^\dfff = \frac{4\log 2}{\pi}\frac{r^2}{\sgI\sgO}$
    &
    $\ds\ell(\om)=\sqrt{\frac{D}{\om}}$\\
    \hline
  \begin{tabular}{c}
    {\bf Acoustic}
    \\
    Speech
  \end{tabular}
    &
    \begin{tabular}{c}
      medium density (kg/m$^3$)
      \vspace*{4pt}
      \\
      $\p_t^2\lm - c^2(\tau \p_t)^\et\nabla^2\lm= c^2\nabla^2\lm$
    \end{tabular}
    &
    \begin{tabular}{c}
      $I(t) =$ injected mass (kg/s)
      \vspace*{4pt}
      \\
      $O(t) =$ excess density (kg)
    \end{tabular}
    &
    $\ds\costfn^\ac = \frac{2\log 2}{\pi}\frac{r^2\ell_\sg^2}{\sgI\sgO^3}$
    &
    \begin{tabular}{c}
      $\ds \nu = (i\tau\om)^\et$
      \\
      $\ds \ell=(c/\om)/\tn{Im}(\nu)$
      \\
      $\ds \ell_\sg=(c/\om)\tn{Im}(\nu)$
    \end{tabular}
    \\
    \hline
  \end{tabular}
  \end{center}
  \caption{
    A summary of the setup and energetic cost to send information for each of the four systems discussed.
    Each system has a coupling field, $\lm$, which is used to transmit a signal between a sender and receiver.
    The dynamics of $\lm$ are given by the listed background dynamics plus the action of the sender $I$ which couples to $\lm$ via $\p_t\lm = (\p_t\lm)_0 + J(x,t)$.
    $\costfn$ is the energetic cost to send information in units of $\kbt/\tn{bit}$ when the transmission distance is below the characteristic lengthscale ($r\ll \ell$).
    {\bf Electrical}: Dynamics are obtained by treating the membrane as a capacitor embedded within a bulk. The limiting lengthscale is given by the RC {\it velocity} $v=\al/c$ considered at frequency $\om$.
    {\bf Diffusion}: Dynamics are standard for a diffusion process. The expected travel distance within a half-period of the signal is $\Delta x = \sqrt{2\pi dD/\om}$, which is the origin of the lengthscale $\ell=\sqrt{D/\om}$.
    {\bf Acoustic}: Dynamics are compressive waves with viscous loss, modeled with fractional Kelvin-Voigt model~\cite{holm_waves_2019} ($\et=1$ for ideal monatomic gases). The lengthscale $\ell$, the inverse attenuation coefficient, limits the transmission distance $r$.
    There is another lengthscale $\ell_\sg$ which limits the listed energetic cost to the regime $\sgI\ll\ell_\sg$, the regime most relevant at the sub-cellular level (see appendix \ref{app:sec:acoustic:parameter-values}).
  }
  \label{tab:results}
\end{table*}

\section{Results}

\begin{figure*}
    \includegraphics[width=0.32\linewidth]{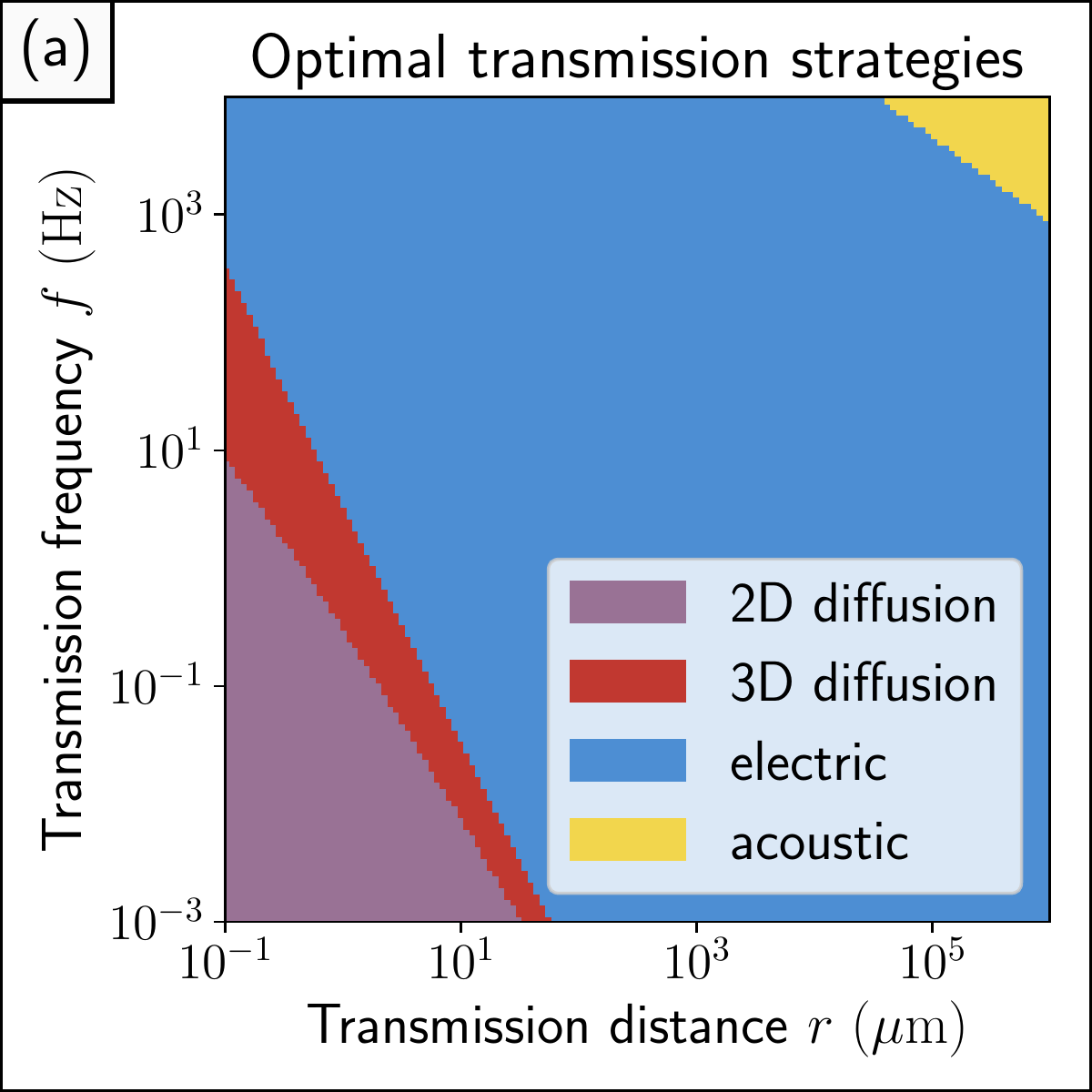}
    \includegraphics[width=0.32\linewidth]{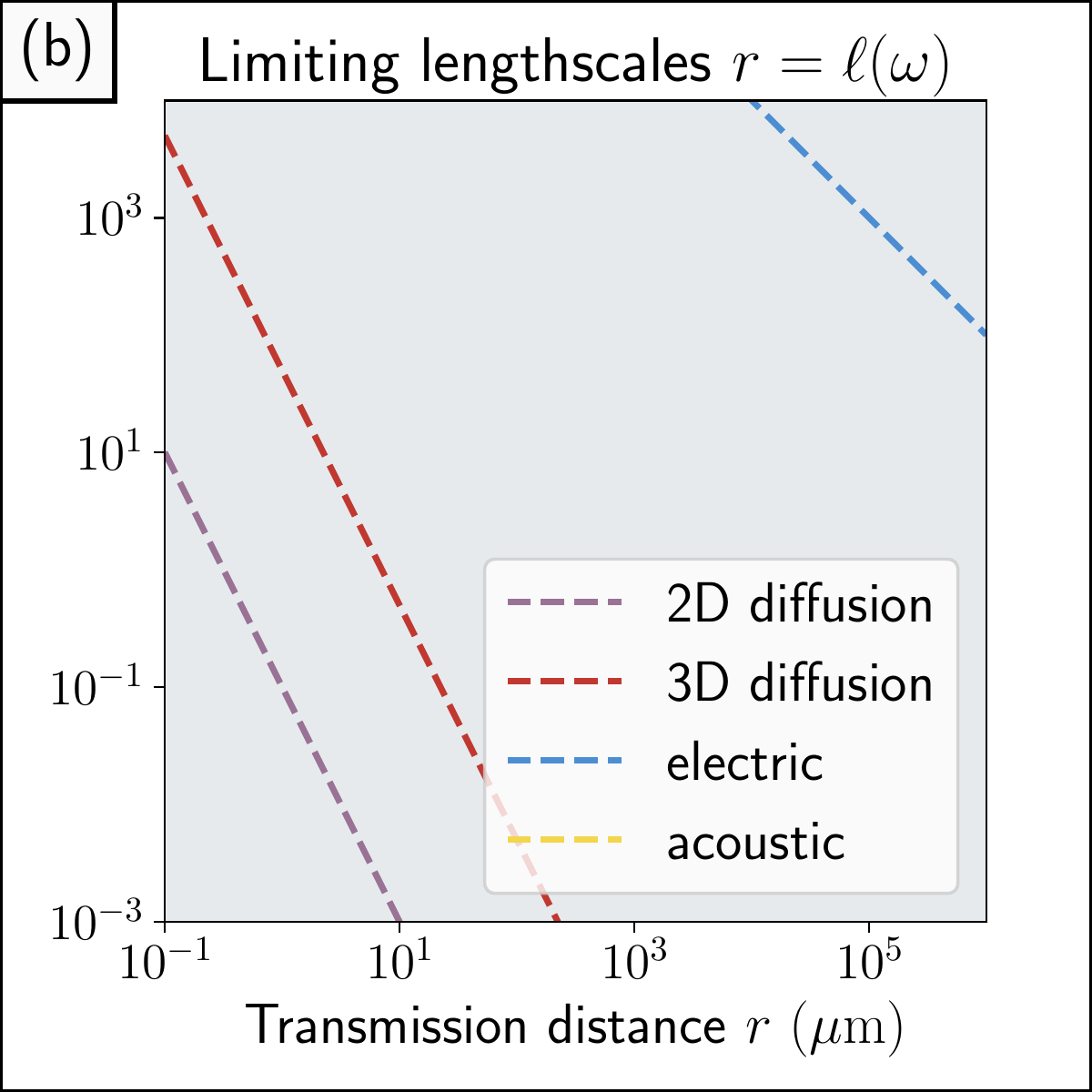}
    \includegraphics[width=0.32\linewidth]{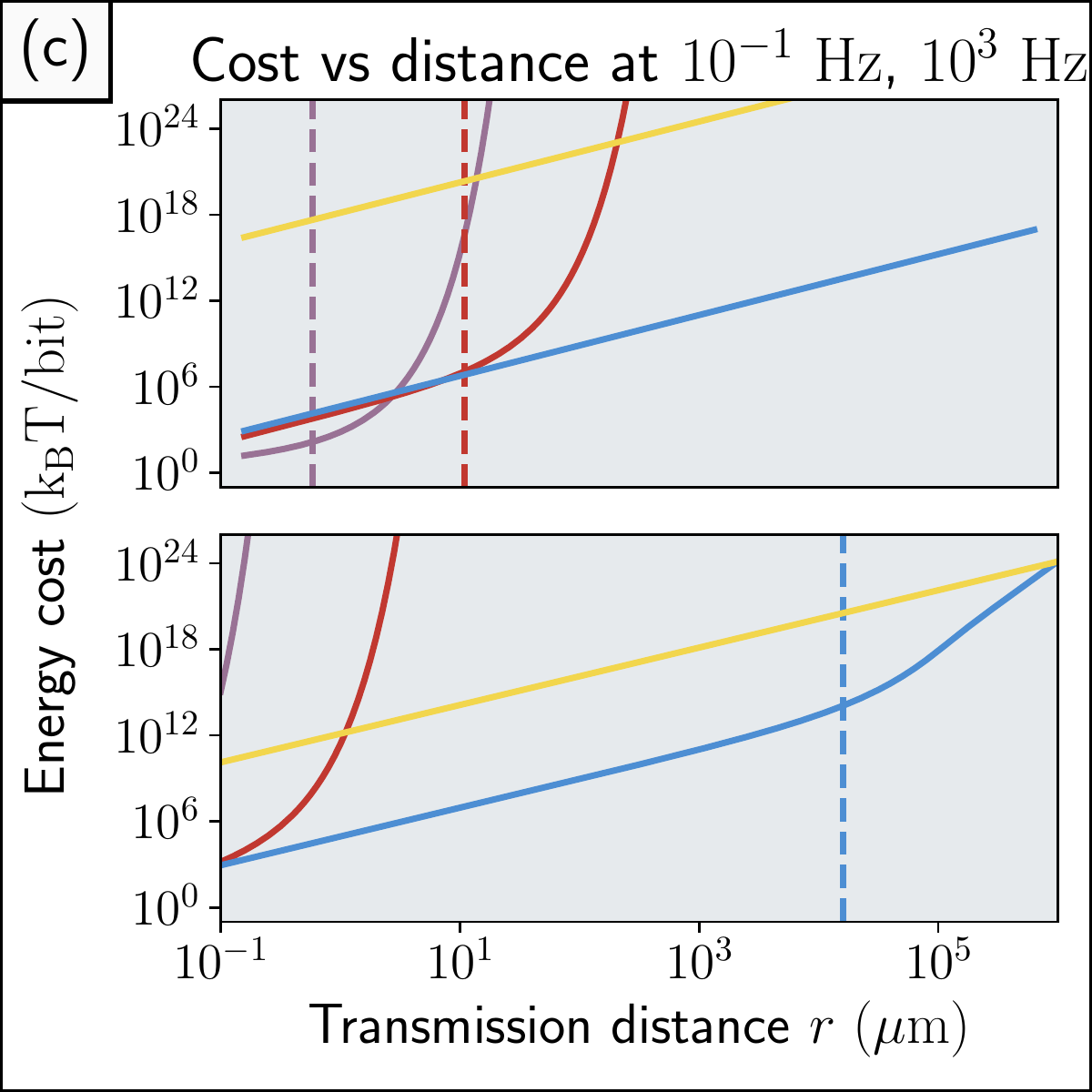}
\caption{\label{fig:phase-diagram}
  An illustration of the energetic cost of signal propagation for four signaling mechanisms.
  2D diffusive signaling shown in purple, 3D diffusive signaling shown in red, electrical-membrane signaling shown in blue, acoustic signaling in saline water shown in yellow.
  For these plots, the sender and receiver sizes are fixed at $\sgI=\sgO=5$ nm. The 3D and 2D diffusion constants are set to $D=50$ $\tn{$\mu$m}^2/\tn{s}$ and $D=0.1$ $\tn{$\mu$m}^2/\tn{s}$ respectively, which are typical values in the cytoplasm and plasma membrane~\cite{schavemaker_how_2018}.
  For electrical signaling, the conductance is $\al=10^{-6}$ S/$\mu\tn{m}$~\cite{wang_role_2013}, the capacitance is $c=10^{-14}$ F/$\mu$m$^2$~\cite{gentet_direct_2000}.
  For acoustic signaling, the wave velocity is $c=1.5\cdot 10^{9}$ $\mu$m/s, the damping parameters $\et=0.21$ and $\tau^\et=8.9\cdot 10^{-5}$ (s$^\et$) which were extracted from acoustic measurements in blood (see appendix \ref{app:sec:acoustic:parameter-values}).
  (a) The optimality phase space. For each value of signal distance and frequency, the color of the optimal signaling mechanism is displayed.
  (b) The characteristic lengthscale $\ell(\om)$ is plotted for each signaling mechanism. These draw exclusion zones dictating where signaling mechanisms become prohibitively expensive.
  (c) The energetic cost of sending information is plotted for each signaling mechanism as a function of distance at a transmission frequency of $0.1$ Hz (top) and $1$ kHz (bottom). The vertical dotted lines correspond to the characteristic lengthscales for each system.
}
\end{figure*}

In the previous section (and methods sections \ref{sec:methods:diffusion}, \ref{sec:methods:acoustic}) we computed the energetic cost of sending information across physical distances through four biologically relevant mediums: electrical signaling, diffusive signaling in 2D and 3D, and acoustic signaling.
Using these calculations we can construct a phase diagram (fig. \ref{fig:phase-diagram}a) with respect to frequency and distance indicating where each method of signaling is energetically preferred (for sensor sizes fixed at $5$ nm).

As summarized in table~\ref{tab:results} (and plotted in fig.~\ref{fig:phase-diagram}b), each of these systems has a characteristic lengthscale $\ell(\om)$ which determines its limits of viability.
When the transmission distance exceeds this lengthscale, the energetic cost of sending information no longer follows the scaling forms $\costfn$, instead becoming exponentially expensive.
Below their characteristic lengthscales, the cost of communication for 3D diffusion, electrical and acoustic signaling scales with distance squared.
Therefore the transition lines between the phases in fig.~\ref{fig:phase-diagram}a can be determined by their characteristic lengthscales, which sets the cutoff in energetic efficiency.
Diffusive signaling in 2D has a unique scaling form and is preferred over 3D diffusive signaling except in the regime where diffusion in 2D would be too slow; typical membrane diffusion constants are 2-3 orders of magnitude slower than those for small molecules in the cytoplasm.

For diffusive and ion-channel signaling, by examining the frequency-dependent characteristic lengthscales, we can see that the frequency $\omega$ can be interpreted as measuring the speed of signal transmission.
These two physical mediums do not support coherent waves and so a signal must reach the receiver before it phase shifts and is strongly attenuated.
In diffusive signaling in $d$ dimensions, for example, this means that the transmission distance is limited to the half-period diffusive travel distance $r^2 < \<\Delta x^2\> = 2dDt=2d\pi D/\om$, explaining the appearance of the lengthscale $\ell(\om)=\sqrt{D/\om}$.
Acoustic signaling is different because it permits coherent waves. Its speed is set by the speed of sound, which is large with respect to biological purposes at sub-cellular scales.
In this case, the characteristic lengthscale $\ell$ (the inverse acoustic attenuation coefficient) is instead a measure of the viscous damping of acoustic waves.

Acoustic and electrical mediums both have characteristic lengthscales which are large relative to biological lengthscales, even at high frequencies (see appendices \ref{app:sec:el:paramters} and \ref{app:sec:acoustic:parameter-values}).
Thus, the regimes in which they are efficient (sub-exponential) overlap significantly.
A comparison of their cost functions ($\costfn^\el\propto r^2/\sgI\sgO$, $\costfn^\ac\propto \ell_\sg^2 r^2/\sgI\sgO^3$, ) shows that electrical signaling is preferred when the receiver size is smaller than the acoustic lengthscale $\ell_\sg=\ell_c\nu''$.
This lengthscale, $\ell_\sg$, depends on the speed of sound $c$, the viscous timescale $\tau$, and the damping power $\et$ within cytoplasm and other biological media.
Based on ultrasound measurements in blood~\cite{holm_waves_2019}, we can estimate that the smallest value of $\ell_\sg$ in the frequency ranges considered is $\ell_\sg = 7$ $\mu$m (appendix \ref{app:sec:acoustic:parameter-values}), which is much larger than a single protein complex~\cite{phillips_physical_2013}.
Thus, at the sub-cellular level where $\sgO\sim 5$ nm, electrical signaling is favored by a factor of $\sim 10^6$.
This provides a possible energetic explanation for why acoustic signaling is absent at the cellular level, despite its omnipresence in the animal kingdom.

\section{Outlook}
The energetic costs we have obtained are lower bounds which hold regardless of the molecular mechanism being used to power the communication channel, which can be quite varied.
For example, diffusive signaling can be driven by futile cycling of reactions (e.g. in CheY phosphorylation/dephosphorylation cycles which in net hydrolyze ATP), or by chemical pumps which concentrate signal molecules for controlled release (e.g. neurotransmitter concentration in synaptic vesicles).
For each of these processes, the total dissipation required to power the channel may be larger than the bounds we derive, but it cannot be smaller.

While an important first step, there are some limitations in this analysis.
Our results often assume a simplified geometry.  In particular, for electrical signaling, many biological examples take place in roughly cylindrical axons and dendrites rather than 2D sheets, qualitatively changing results when signals are sent a distance farther than the cylinder radius.
Our setup also doesn't capture some physical strategies such as the manipulation of stresses in fiber networks and directed transport by motor proteins.
Lastly, the noise in this calculation is assumed to be thermal noise arising from equilibrium fluctuations.
Equilibrium provides a well-defined statistical structure permitting a concrete information theoretic analysis.
However, biology does not operate at equilibrium.
While the result is still a lower bound, a higher bound could be obtained by characterizing the statistical features relevant to the process being studied.

There may also be other design principles which are not covered by our bounds.
We do not consider the cost of building and maintaining the protein machinery required to run these communication channels.
Certain strategies also permit multiplexing. For example, in diffusive signaling, multiple signal networks can run in parallel by using different second messenger molecules.
In contrast, there is only one electrical potential.

For sending information over longer ranges biology often uses relays, where information is sent through an excitable medium, often as a traveling wave, a mechanism of information transfer not discussed in this work.
For example, while action potentials are certainly electrical, the signal is not sent from the cell body all the way to the end of an axon directly, but instead via a relay of ion channels triggering successive channels.
However, it is possible our approach may be extended to consider signaling in excitable media, but the present analysis does not consider this.

The driving motivation behind this work is to understand the energetic cost of biological computations constrained by the physical reality of biological environments.
In conjunction with a range of recent efforts to quantify information transfer across biological scales \cite{Nemenman04,Gregor07,Mattingly21} we hope that follow ups to this work will be able to quantify a computational budget required for these processes, which in many cases appears to be substantial \cite{rodenfels_heat_2019,Attwell01}.

Prior work which investigates the cost of computation often considers either the Landauer limit \cite{landauer_irreversibility_1961,fuchs_stochastic_2016,mehta_energetic_2012,sartori_thermodynamic_2014} required to erase information or the cost of breaking time-reversal symmetry~\cite{feng_length_2008,parrondo_entropy_2009,cao_free-energy_2015,barato_cost_2016,zhang_energy_2019,bryant_energy_2020}.
While these results are fundamental, they produce results on the order of $\kbt$, far below the energetic scale seen in real processes.
In contrast, the results obtained here depend on physical constants as well as practical constraints like the size of the sensors and the transmission distance, producing energetic constraints orders of magnitude larger.
For example, for a diffusive signal sent in 3D over a distance $r\sim 1$ $\mu$m, with sensor sizes on the order of $\sigma\sim 10$ nm, the cost is on the order of $10^4$ $\kbt$.
Thus the large costs that biology must pay to process information \textit{can} be understood theoretically, albeit only by additionally considering the physical constraints on real biological systems.

\section{Acknowledgements}
We thank Isabella Graf, Henry Mattingly, and Mason Rouches for useful comments on the manuscript as well as Sverre Holm for discussions on acoustic phenomena, but especially James Sethna for inspiring discussions. The work was funded by NIH R35 GM138341, NSF 1808551 and a Simons Investigator award.

\section{Methods}
\label{sec:methods}

Here we discuss some some of the missing details from section \ref{sec:cost_per_bit} as well as the derivation done in \ref{sec:electrical} for acoustic and diffusive signaling systems.
Namely we discuss the mutual information rate (eq. (\ref{eq:info_rate}), the energetic cost per bit (eq. (\ref{eq:cost_per_bit})), and the signal/noise/dissipation for the remaining systems.
A thorough derivation for each system can be found in appendices \ref{app:sec:electrical}-\ref{app:sec:acoustic}.

\subsection{Time-series mutual information rate and cost per bit}
\label{sec:methods:cost-per-bit}

We work with the framework used by Tostevin and ten Wolde in ref.~\cite{tostevin_mutual_2009}, which requires us to adapt some of our previous notation.
In the main text, we denoted $S^O(\om)$ to mean the intrinsic noise in the output signal arising from equilibrium fluctuation in the absence of an input signal.
Here we have to be more precise.
We now denote $S^{O}_0(\om)$ (i.e. the noise, $N(\om)$ in~\cite{tostevin_mutual_2009}) to mean the intrinsic noise in $O(t)$.
We denote $S^{O}_I(\om)$ (i.e. the power, $P(\om)$ in ~\cite{tostevin_mutual_2009}) to mean the contribution to the power spectrum in $O(t)$ arising as a result of the input signal. And we denote $S^{O}_T(\om)$ to mean their combined contributions.
As in ref~\cite{tostevin_mutual_2009}, the spectral addition rule holds:
\begin{equation}
  S^{O}_T(\om) = S^{O}_0(\om) + S^{O}_I(\om)
\end{equation}
In the linear response regime, the output of the system can be written in terms of a response function and the input:
\begin{equation}
  \<O(\om)\> = \ch^{OI}(\om)\, I(\om)
\end{equation}
From this, we can rewrite $S^{O}_I(\om)$ in terms of the response kernel:
\begin{equation}
  S^{O}_I(\om) = |\ch^{OI}(\om)|^2S^{I}(\om)
\end{equation}
where $S^{I}(\om)$ is the power spectrum of the input signal.
The mutual information rate between $I$ and $O$ can then be written:
\begin{equation}
  R(I, O) = \frac{1}{4\pi}\int\d\om\ln\l(
    1
    +
    \frac{|\ch^{OI}(\om)|^2S^{I}(\om)}{S_0^O(\om)}
  \r)
\end{equation}
In the weak signal regime, which is a key assumption of this analysis, we can reasonably approximate the previous expression using the normal $\log$ expansion:
\begin{equation}
  \label{eq:methods:mi_rate}
  R(I, O) = \frac{1}{4\pi\log 2}\int\d\om
    \frac{|\ch^{OI}(\om)|^2S^{I}(\om)}{S_0^O(\om)}
  \quad
  (\tn{bits/s})
\end{equation}
where the factor of $\log 2$ converts from nats to bits.
This is eq. (1) from the main text.

In our model, each physical is characterized by its intrinsic noise $S^O_0(\om)$, describing the output's behavior when no signal is present, and its gain $\ch^{OI}(\om)$ which describes how the output signal responds to the input signal.
The specific signal $I(t)$ which the sender chooses to send is not a property of the channel itself.
Since we care about the dissipation incurred by the physical channel used to manifest the network, we also characterize each network by its dissipation kernel:
\begin{equation}
  \tn{Dissipation rate} = \int\d\om\,\dissfn(\om)S^{I}(\om)
\end{equation}
which specifies the total energy dissipation rate required to run a process $I(t)$ in terms of the cost at each frequency.

For a generalized input spectrum $S^{I}(\om)$, the cost to send information is the dissipation rate divided by the information rate:
\begin{equation}
  \label{eq:methods:cost_per_bit_full}
  \hat{\costfn}
  =
    4\pi\log 2\frac{\int\d\om\,\dissfn(\om)S^{I}(\om)}
    {\int\d\om\,|\ch^{OI}(\om)|^2S^{I}(\om)/S_0^O(\om)}
\end{equation}
The energy cost of sending information can be more concisely written in terms of a cost per bit at a particular frequency $\om$,
\begin{equation}
  \label{eq:methods:cost_per_bit}
  \costfn(\om) = 4\pi\log 2
    \frac{\dissfn(\om) S_0^O(\om)}{|\ch^{OI}(\om)|^2},
\end{equation}
which is the form given by eq. (2) in the main text.
This cost can be interpreted as a cost per bit for signals localized in frequency (e.g. a radio signal).
Alternatively, for a signal with power at many frequencies, the total power can be written as an integral over frequency space of $\costfn(\om)$ multiplying the differential information content at that frequency.

\subsection{Diffusive signaling}
\label{sec:methods:diffusion}
Here we present a sketch of the calculation performed in section \ref{sec:electrical} for diffusive signaling in $d$ dimensions.
A thorough derivation can be found in appendix \ref{app:sec:diffusion}.

We suppose there is a sender $I$ and receiver $O$ separated by a distance $r$ as depicted in fig.~\ref{fig:fig1}d.
The fluctuation in particle density is given by $\lm(\bb x,t)=\rh(\bb x,t)-\rh_0$ and is subject to an equilibrium diffusive process $\p_t\lm = D\div^2\lm$.
In a half-period, the distance traveled due to diffusion is $\Delta x^2 = 2dD(T/2)=2\pi dD/\om$.
Thus we can characterize this system by lengthscale $\ell(\om)=\sqrt{D/\om}$.

Analogously to electrical signaling, the sender creates/destroys particles with localized rate $J(\bb x,t)$, and the receiver measures the localized particle excess $O(t)$ (fig.~\ref{fig:fig1}a).
The free energy is purely entropic and is given by density (up to a constant):
\begin{equation}
  \bt \frac{\d F}{\d V} = (\lm(\bb x)+\rh_0)\log\l(1+\frac{\lm(\bb x)}{\rh_0}\r) - \bt\lm(\bb x) h(\bb x)
\end{equation}
Where $h$ is an artificial external field which couples to $\lm$ in the free energy, introduced for the purpose of computing the susceptibility and then the noise.
From this, we can derive chemical potential $\mu=\dl F/\dl\rh = \kbt\l(1 + \log(1 + \lm/\rh_0\r) - h(\bb x)$, and potential current $J_\mu=-\bt D\rh\div\mu$ which is subject to the conservation equation: $\p_t\lm = - \div J_\mu$.
Solving this yields the dynamics for the system:
\begin{equation}
  \p_t\lm = D\div^2(\lm-\bt\rh_0 h) - D\bt\l[(\div\lm)(\div h) + \lm\div^2 h\r]+J
\end{equation}
where we have added the contribution $J$ from the sender.
By Fourier transforming to momentum-frequency space and linearizing the dynamics, we obtain response functions:
\begin{align}
  \ch^h = \frac{\bt\rh_0k^2}{k^2+i\om/D}
  &&
  \ch^J = \frac{1}{Dk^2+i\om}
\end{align}
Then following the same logic as section \ref{sec:electrical}, we can use $\ch^h$ to compute the noise $S^O$:
\begin{align}
  \label{eq:df:noise}
  \begin{split}
    \tn{($d=3$)}
    \qquad\quad
    S^O(\om) &= \frac{4\pi^{3/2}\rh_0\sgO^5}{D}U^\dfff_N(\sgO/\ell(\om))
    \\
    \tn{($d=2$)}
    \qquad\quad
    S^O(\om) &= \frac{2\pi\rh_0\sgO^4}{D}U^\dff_N(\sgO/\ell(\om))
  \end{split}
\end{align}
with scaling behavior $U^\dfff_N(y\to 0)\sim 1$, $U^\dff_N(y\to 0)\sim -2\log(y)$, with the latter diverging as $y\to 0$.
This divergence is why the scaling form of the energy, $\costfn^\dff$, depends explicitly on $\ell(\om)$.
We use $\ch^I$ to compute the gain $|\ch^{OI}|^2$:
\begin{equation}
  \label{eq:df:signal}
  \begin{split}
    \tn{($d=3$)}
    \qquad\quad
    \l|\ch^{OI}(\om)\r|^2 &= \frac{\pi}{2}\frac{\sgO^6}{D^2r^2}U^\dfff_S(r/\ell(\om))
    \\
    \tn{($d=2$)}
    \qquad\quad
    \l|\ch^{OI}(\om)\r|^2 &= \frac{\sgO^4}{D^2}U^\dff_S(r/\ell(\om))
  \end{split}
\end{equation}
with scaling behavior $U^\dfff_S(z\to 0)\sim 1$, $U^\dff_S(z\to 0)\sim \log^2(r/\ell)$.
Finally we can compute the dissipation by looking at the change in free energy induced by the signal $\<\tn{diss}\>=\l\<\frac{\p F}{\p\lm}J\r\>_t$. This yields a dissipation kernel:
\begin{equation}
  \label{eq:df:diss}
  \begin{split}
    \tn{($d=3$)}
    \qquad\quad
    \dissfn(\om) &= \frac{\kbt}{8\pi^{5/2}\rh_0D\sgI}U^\dfff_D\l(\frac{\sgI}{\ell(\om)}\r)
    \\
    \tn{($d=2$)}
    \qquad\quad
    \dissfn(\om) &= \frac{\kbt}{8\pi^2\rh_0D}U^\dff_D\l(\frac{\sgI}{\ell(\om)}\r)
  \end{split}
\end{equation}
with scaling behavior $U^\dfff_D(x\to 0)\sim 1$, $U^\dff_D(x\to 0)\to -2\log(x)$.
Then plugging equations (\ref{eq:df:noise}-\ref{eq:df:diss}) into the energetic cost (\ref{eq:cost_per_bit}), we find the energetic cost of sending a diffusive signal over a distance $r$ in units of $\kbt/$bit to be:
\begin{equation}
  \label{eq:df:result}
  \begin{split}
    \tn{($d=3$)}
    \qquad\quad
    \costfn^\dfff
    &=
    \frac{4\log 2}{\pi}\frac{r^2}{\sgI\sgO}U^\dfff\l(\frac{r}{\ell(\om)}\r)
    \\
    \tn{($d=3$)}
    \qquad\quad
    \costfn^\dff
    &\approx 4\log 2\frac{\log\l(\frac{\ell(\om)}{\sgI}\r)\log\l(\frac{\ell(\om)}{\sgO}\r)}{\log\l(\frac{\ell(\om)}{r}\r)^2}
  \end{split}
\end{equation}
we note that $U^\dfff(r/\ell\to 0)\to 1$ is a small correction to the scaling form of $\costfn^\dfff$.
Diffusion in 2D behaves differently from all the other systems because the scaling form is given entirely by the asymptotic behavior of the integral functions $U^\dff_N$, $U^\dff_S$, $U^\dff_D$.
For a discussion of these asymptotic relations, see appendices \ref{app:sec:diffusion:signal}-\ref{app:sec:diffusion:dissipation}.

\subsection{Acoustic signaling}
\label{sec:methods:acoustic}
Here we present a sketch of the calculation performed in section \ref{sec:electrical} for acoustic signaling in 3 dimensions in a liquid.
A thorough derivation can be found in appendix \ref{app:sec:acoustic}.

There is a sender $I$ at the origin and a receiver $O$ at a point $\bb r$ embedded in a 3D media which supports coherent compression waves with viscous loss.
A physical wave may be characterized by its displacement vector $\bb u(\bb x,t)$.
We use the fractional Kelvin-Voigt model to describe the background dynamics of $\bb u$ which is the standard wave-equation with a fractional viscous loss term~\cite{wismer_finite_2006,holm_unifying_2010,holm_waves_2019}:
\begin{equation}
  \label{eq:methods:ac-kelvin-voight}
  \p_t^2\bb u - c^2\div^2\bb u - c^2\l[\tau \p_t\r]^\et\div^2\bb u = 0
\end{equation}
where $c$ is the speed of sound, $\tau$ is a timescale characterizing the strength of the relaxation modes, $0<\et\leq 1$ is a phenomenological fractional power reflecting the relaxation dependence on frequency, and $\p_t^\et$ is a fractional derivative, which for our purposes is defined by its Fourier transform:
  $\mathcal{F}\l\{\frac{\p^\et}{\p t^\et}f\r\}(\om)
  =
  (i\om)^\et
  \mathcal{F}\l\{f\r\}(\om)$.
This is an accurate model for low-frequency acoustics in liquids and gases~\cite{morse_theoretical_1968,holm_waves_2019}.
For ideal monatomic gases or distilled water, $\et=1$, which leads to characteristic lengthscale $\ell=c/\om^2\tau$, the classical $\om^2$ attenuation discovered by Stokes~\cite{stokes_theories_2009}.
We characterize this system in terms of the complex damping parameter:
\begin{equation}
  \label{eq:methods:ac-nu}
  \nu = \nu' + i\nu'' \equiv (i\tau\om)^\et
\end{equation}
and a few (over-complete) lengthscales:
\begin{align}
  \label{eq:ac:parameters}
  \ell_c \equiv \frac{c}{\om}
  &&
  \ell = \frac{\ell_c}{\nu''}
  &&
  \ell_\sg\equiv \ell_c\nu''
\end{align}
where $\ell_c$ is the radian wavelength and $\ell$ is more commonly expressed as $\al=1/2\ell$, the acoustic attenuation coefficient~\cite{morse_theoretical_1968,dukhin_bulk_2009,holmes_temperature_2011,lin_bulk_2017,holm_waves_2019}.
The lengthscale $\ell_\sg$ pops out of our analysis as a limiting lengthscale for the sender size $\sgI$ as well as term that appears in the final energetic cost.

We will describe the dynamics in terms of the density fluctuations $\lm = -\rh_0\div\cdot\bb u$ (equivalently we could use the pressure fluctuations).
In this framing, the input signal $J$ and output $O(t)$ take on the form depicted in fig.~\ref{fig:fig1}a.
The energy density is given by~\cite{morse_theoretical_1968}:
\begin{equation}
  \frac{\d H}{\d V} = \frac{1}{2}(\rh_0+\lm(x,t))\l(\frac{\p\bb u}{\p t}\r)^2 + \frac{(p_0 + c^2\lm)^2}{2\rh_0 c^2} + h\lm
\end{equation}
where $p_0$ is the equilibrium pressure and $h$ is an artificial external field introduced to compute the susceptibility.
Solving the Euler-Lagrange equations then inserting the viscous loss term and signal term yields the full dynamics:
\begin{equation}
  (\p_t^2 - c^2\div^2- c^2(\tau\p_t)^\et\div^2)\lm = \rh_0\div^2 h + \p_t J
\end{equation}
from which we can extract the response functions:
\begin{align}
  \ch^h = \frac{(\rh_0/c^2) k^2}{k^2-\ell_c^{-2}+\nu k^2}
  &&
  \ch^J = \frac{i}{\om}\frac{\ell_c^{-2}}{k^2-\ell_c^{-2}+\nu k^2}
\end{align}
Then following through the calculation in section \ref{sec:electrical} we use $\ch^h$ to obtain the noise $S^O(\om)$ via the fluctuation dissipation theorem~\cite{tong_lectures_2012}:
\begin{equation}
  \label{eq:ac:noise}
  S^O(\om) = 2\pi^{3/2}\frac{\rh_0}{\bt}\frac{\sgO^3\nu''}{c^2\om}U^\ac_N\l(\frac{\sgO}{\ell_c}, \nu\r)
\end{equation}
We discuss the scaling of $U^\ac_N$ after equation (\ref{eq:ac:result}).
We obtain the gain $\l|\ch^{OI}\r|^2$ using $\ch^{J}$:
\begin{equation}
  \label{eq:ac:signal}
  \l|\ch^{OI}\r|^2 = \frac{\pi}{2}\l(\frac{\om\sgO^3}{rc^2}\r)^2U^\ac_S\l(\frac{\sgO}{\ell_c}, \nu\r)
\end{equation}
with scaling $U^\ac_S(z\to 0, \nu\to 0)\sim 1$.
We can compute the dissipation by looking at the linearized change (i.e. ignoring the kinetic term) in energy induced by the signal: $\<\tn{diss}\> = \<\frac{\p H}{\p\lm}J\>$, which yields dissipation kernel:
\begin{equation}
  \label{eq:ac:diss}
  \dissfn(\om) = \frac{1}{8\pi^{5/2}\rh_0}\frac{\om\nu''}{\sgI}U^\ac_D\l(\frac{\sgO}{\ell_c}, \nu\r)
\end{equation}
Plugging equations (\ref{eq:ac:noise}-\ref{eq:ac:diss}) into the energetic cost (\ref{eq:cost_per_bit}), we find the energetic cost of sending an acoustic signal over a distance $r$ in units of $\kbt/$bit to be:
\begin{equation}
  \label{eq:ac:result}
  \costfn^\ac = \frac{2\log 2}{\pi}\frac{r^2\ell_\sg^2}{\sgI\sgO^3}
    \frac{U^\ac_N\l(\frac{\sgO}{\ell_c}, \nu\r)U^\ac_D\l(\frac{\sgO}{\ell_c}, \nu\r)}
      {U^\ac_S\l(\frac{\sgO}{\ell_c}, \nu\r)}
\end{equation}
where $\ell_\sg=\ell_c\nu''$.
The asymptotic behavior of $U^\ac_N$, $U^\ac_D$, and $U^\ac_S$ for small arguments depends on the ratio of their first and second argument and is developed explicitly in appendices \ref{app:sec:acoustic:signal}-\ref{app:sec:acoustic:dissipation}.
The result is that in the regime where the characteristic damping is small ($\nu\ll 1$)
and where the sender and receiver sizes are small compared to $\ell_\sg$, then the noise and dissipation functions may be neglected ($U^\ac_N\to 1$, $U^\ac_D\to 1$) and the signal function is given by $U^\ac_S(r/\ell_c, \nu) \sim e^{-r/\ell}$.
This latter scaling relation is commonly recognized as the defining feature of the attenuation coefficient $\al=1/2\ell$~\cite{morse_theoretical_1968}.

Whether we are in the regime $\sgI$, $\sgO\ll\ell_\sg$ depends on the damping parameters $\tau$ and $\et$ which are, in general, difficult to compute.
For distilled water, $\et=1$ and we can compute the damping timescale using $\tau=(1/\rh_0c^2)\l[4\et/3 + \mu_B\r] = 1.6\cdot 10^{-12}$ s where $\et$ and $\et_B$ are the dynamic and bulk viscosities~\cite{holm_waves_2019,holmes_temperature_2011}. This would mean that the scaling regime is only valid when $\sgI\ll \tau_\ell=2.4$ nm.
However, biology does not operate in pure water.
As a proxy for relevant biological fluids, we can extract $\et$ and $\tau$ from measurements of the acoustic attenuation coefficient in blood studied in the context of ultrasound experiments (Supplementary section 4.2).
The result of this procedure is that $\ell_\sg = 7$ ($\mu$m) at the largest frequency shown in fig.~\ref{fig:phase-diagram}a and becomes even larger for smaller frequencies, and thus we may assume sub-cellular senders/receivers to be significantly smaller.

\bibliography{paper_bibliography}

\begin{thebibliography}{10}

\bibitem{hopfield_kinetic_1974}
J.~J. Hopfield, ``Kinetic {Proofreading}: {A} {New} {Mechanism} for {Reducing}
  {Errors} in {Biosynthetic} {Processes} {Requiring} {High} {Specificity},''
  {\em Proceedings of the National Academy of Sciences}, vol.~71,
  pp.~4135--4139, Oct. 1974.

\bibitem{landauer_irreversibility_1961}
R.~Landauer, ``Irreversibility and {Heat} {Generation} in the {Computing}
  {Process},'' {\em IBM Journal of Research and Development}, vol.~5,
  pp.~183--191, July 1961.

\bibitem{bennett_thermodynamics_1982}
C.~H. Bennett, ``The thermodynamics of computation—a review,'' {\em
  International Journal of Theoretical Physics}, vol.~21, pp.~905--940, Dec.
  1982.

\bibitem{wolpert_stochastic_2019}
D.~H. Wolpert, ``The stochastic thermodynamics of computation,'' {\em Journal
  of Physics A: Mathematical and Theoretical}, vol.~52, p.~193001, May 2019.

\bibitem{chen_deterministic_2013}
H.-L. Chen, D.~Doty, and D.~Soloveichik, ``Deterministic {Function}
  {Computation} with {Chemical} {Reaction} {Networks},'' {\em arXiv:1204.4176
  [cs]}, Jan. 2013.

\bibitem{kolchinsky_entropy_2021}
A.~Kolchinsky and D.~H. Wolpert, ``Entropy production given constraints on the
  energy functions,'' {\em Physical Review E}, vol.~104, p.~034129, Sept. 2021.

\bibitem{mehta_energetic_2012}
P.~Mehta and D.~J. Schwab, ``Energetic costs of cellular computation,'' {\em
  Proceedings of the National Academy of Sciences}, vol.~109, pp.~17978--17982,
  Oct. 2012.

\bibitem{govern_energy_2014}
C.~C. Govern and P.~R.~T. Wolde, ``Energy dissipation and noise correlations in
  biochemical sensing,'' {\em Physical Review Letters}, vol.~113, no.~25,
  pp.~undefined--undefined, 2014.

\bibitem{wang_price_2020}
T.-L. Wang, B.~Kuznets-Speck, J.~Broderick, and M.~Hinczewski, ``The price of a
  bit: energetic costs and the evolution of cellular signaling,'' {\em
  bioRxiv}, p.~2020.10.06.327700, Oct. 2020.

\bibitem{ouldridge_thermodynamics_2017}
T.~E. Ouldridge, C.~C. Govern, and P.~R. ten Wolde, ``Thermodynamics of
  {Computational} {Copying} in {Biochemical} {Systems},'' {\em Physical Review
  X}, vol.~7, p.~021004, Apr. 2017.

\bibitem{sartori_thermodynamic_2014}
P.~Sartori, L.~Granger, C.~F. Lee, and J.~M. Horowitz, ``Thermodynamic {Costs}
  of {Information} {Processing} in {Sensory} {Adaptation},'' {\em PLoS
  Computational Biology}, vol.~10, p.~e1003974, Dec. 2014.

\bibitem{barato_information-theoretic_2013}
A.~C. Barato, D.~Hartich, and U.~Seifert, ``Information-theoretic vs.
  thermodynamic entropy production in autonomous sensory networks,'' {\em
  Physical Review E}, vol.~87, Apr. 2013.

\bibitem{feng_length_2008}
E.~H. Feng and G.~E. Crooks, ``Length of {Time}’s {Arrow},'' {\em Physical
  Review Letters}, vol.~101, Aug. 2008.

\bibitem{parrondo_entropy_2009}
J.~M.~R. Parrondo, C.~V.~d. Broeck, and R.~Kawai, ``Entropy production and the
  arrow of time,'' {\em New Journal of Physics}, vol.~11, p.~073008, July 2009.

\bibitem{barato_thermodynamic_2015}
A.~C. Barato and U.~Seifert, ``Thermodynamic {Uncertainty} {Relation} for
  {Biomolecular} {Processes},'' {\em Physical Review Letters}, vol.~114, Apr.
  2015.

\bibitem{brown_effective_2016}
A.~I. Brown and D.~A. Sivak, ``Effective dissipation: {Breaking} time-reversal
  symmetry in driven microscopic energy transmission,'' {\em Physical Review
  E}, vol.~94, Sept. 2016.

\bibitem{rao_nonequilibrium_2016}
R.~Rao and M.~Esposito, ``Nonequilibrium {Thermodynamics} of {Chemical}
  {Reaction} {Networks}: {Wisdom} from {Stochastic} {Thermodynamics},'' {\em
  Physical Review X}, vol.~6, p.~041064, Dec. 2016.

\bibitem{cao_free-energy_2015}
Y.~Cao, H.~Wang, Q.~Ouyang, and Y.~Tu, ``The free-energy cost of accurate
  biochemical oscillations,'' {\em Nature Physics}, vol.~11, pp.~772--778,
  Sept. 2015.

\bibitem{barato_cost_2016}
A.~C. Barato and U.~Seifert, ``Cost and {Precision} of {Brownian} {Clocks},''
  {\em Physical Review X}, vol.~6, Dec. 2016.

\bibitem{zhang_energy_2019}
D.~Zhang, Y.~Cao, Q.~Ouyang, and Y.~Tu, ``The energy cost and optimal design
  for synchronization of coupled molecular oscillators,'' {\em Nature Physics},
  Nov. 2019.

\bibitem{england_statistical_2013}
J.~L. England, ``Statistical physics of self-replication,'' {\em The Journal of
  Chemical Physics}, vol.~139, p.~121923, Sept. 2013.

\bibitem{sivak_thermodynamic_2012}
D.~A. Sivak and G.~E. Crooks, ``Thermodynamic metrics and optimal paths,'' {\em
  Physical Review Letters}, vol.~108, May 2012.

\bibitem{Machta15}
B.~B. Machta, ``Dissipation bound for thermodynamic control,'' {\em Phys. Rev.
  Lett.}, vol.~115, p.~260603, Dec 2015.

\bibitem{bryant_energy_2020}
S.~J. Bryant and B.~B. Machta, ``Energy dissipation bounds for autonomous
  thermodynamic cycles,'' {\em Proceedings of the National Academy of
  Sciences}, vol.~117, pp.~3478--3483, Feb. 2020.

\bibitem{rodenfels_heat_2019}
J.~Rodenfels, K.~M. Neugebauer, and J.~Howard, ``Heat {Oscillations} {Driven}
  by the {Embryonic} {Cell} {Cycle} {Reveal} the {Energetic} {Costs} of
  {Signaling},'' {\em Developmental Cell}, vol.~48, pp.~646--658.e6, Mar. 2019.

\bibitem{Levy21}
W.~B. Levy and V.~G. Calvert, ``Communication consumes 35 times more energy
  than computation in the human cortex, but both costs are needed to predict
  synapse number,'' {\em Proceedings of the National Academy of Sciences},
  vol.~118, no.~18, p.~e2008173118, 2021.

\bibitem{Attwell01}
D.~Attwell and S.~B. Laughlin, ``An energy budget for signaling in the grey
  matter of the brain,'' {\em Journal of Cerebral Blood Flow \& Metabolism},
  vol.~21, no.~10, pp.~1133--1145, 2001.

\bibitem{Laughlin98}
S.~B. Laughlin, R.~R. de~Ruyter~van Steveninck, and J.~C. Anderson, ``The
  metabolic cost of neural information,'' {\em Nature Neuroscience}, vol.~1,
  pp.~36--41, May 1998.

\bibitem{smith_cerebral_2002}
A.~J. Smith, H.~Blumenfeld, K.~L. Behar, D.~L. Rothman, R.~G. Shulman, and
  F.~Hyder, ``Cerebral energetics and spiking frequency: {The}
  neurophysiological basis of {fMRI},'' {\em Proceedings of the National
  Academy of Sciences}, vol.~99, pp.~10765--10770, Aug. 2002.

\bibitem{raichle_appraising_2002}
M.~E. Raichle and D.~A. Gusnard, ``Appraising the brain's energy budget,'' {\em
  Proceedings of the National Academy of Sciences}, vol.~99, pp.~10237--10239,
  Aug. 2002.

\bibitem{tong_lectures_2012}
D.~Tong, ``Lectures on {Kinetic} {Theory},'' University of Cambridge, 2012.

\bibitem{tostevin_mutual_2009}
F.~Tostevin and P.~R. ten Wolde, ``Mutual {Information} between {Input} and
  {Output} {Trajectories} of {Biochemical} {Networks},'' {\em Physical Review
  Letters}, vol.~102, May 2009.

\bibitem{zwanzig_nonequilibrium_2001}
R.~Zwanzig, {\em Nonequilibrium statistical mechanics}.
\newblock Oxford ; New York: Oxford University Press, 2001.

\bibitem{holm_waves_2019}
S.~Holm, {\em Waves with {Power}-{Law} {Attenuation}}.
\newblock Cham: Springer International Publishing, 2019.

\bibitem{phillips_physical_2013}
R.~Phillips, {\em Physical biology of the cell}.
\newblock London : New York, NY: Garland Science, second edition~ed., 2013.

\bibitem{Nemenman04}
I.~Nemenman, W.~Bialek, and R.~de~Ruyter~van Steveninck, ``Entropy and
  information in neural spike trains: Progress on the sampling problem,'' {\em
  Phys. Rev. E}, vol.~69, p.~056111, May 2004.

\bibitem{Gregor07}
T.~Gregor, D.~W. Tank, E.~F. Wieschaus, and W.~Bialek, ``Probing the {Limits}
  to {Positional} {Information},'' {\em Cell}, vol.~130, no.~1, pp.~153--164,
  2007.

\bibitem{Mattingly21}
H.~H. Mattingly, K.~Kamino, B.~B. Machta, and T.~Emonet, ``Escherichia coli
  chemotaxis is information limited,'' {\em Nature Physics}, vol.~17,
  pp.~1426--1431, Dec. 2021.

\bibitem{fuchs_stochastic_2016}
J.~Fuchs, S.~Goldt, and U.~Seifert, ``Stochastic thermodynamics of resetting,''
  {\em EPL (Europhysics Letters)}, vol.~113, p.~60009, Mar. 2016.

\bibitem{schavemaker_how_2018}
P.~E. Schavemaker, A.~J. Boersma, and B.~Poolman, ``How {Important} {Is}
  {Protein} {Diffusion} in {Prokaryotes}?,'' {\em Frontiers in Molecular
  Biosciences}, vol.~5, 2018.

\bibitem{wang_role_2013}
K.~Wang, J.~Riera, H.~Enjieu-Kadji, and R.~Kawashima, ``The {Role} of
  {Extracellular} {Conductivity} {Profiles} in {Compartmental} {Models} for
  {Neurons}: {Particulars} for {Layer} 5 {Pyramidal} {Cells},'' {\em Neural
  Computation}, vol.~25, pp.~1807--1852, July 2013.

\bibitem{gentet_direct_2000}
L.~J. Gentet, G.~J. Stuart, and J.~D. Clements, ``Direct measurement of
  specific membrane capacitance in neurons.,'' {\em Biophysical Journal},
  vol.~79, pp.~314--320, July 2000.

\bibitem{wismer_finite_2006}
M.~G. Wismer, ``Finite element analysis of broadband acoustic pulses through
  inhomogenous media with power law attenuation,'' {\em The Journal of the
  Acoustical Society of America}, vol.~120, pp.~3493--3502, Dec. 2006.

\bibitem{holm_unifying_2010}
S.~Holm and R.~Sinkus, ``A unifying fractional wave equation for compressional
  and shear waves,'' {\em The Journal of the Acoustical Society of America},
  vol.~127, pp.~542--548, Jan. 2010.

\bibitem{morse_theoretical_1968}
P.~M. Morse and K.~U. Ingard, {\em Theoretical {Acoustics}}.
\newblock McGraw-Hill, 1968.

\bibitem{stokes_theories_2009}
``On the {Theories} of the {Internal} {Friction} of {Fluids} in {Motion}, and
  of the {Equilibrium} and {Motion} of {Elastic} {Solids},'' in {\em
  Mathematical and {Physical} {Papers}} (G.~G. Stokes, ed.), vol.~1 of {\em
  Cambridge {Library} {Collection} - {Mathematics}}, pp.~75--129, Cambridge:
  Cambridge University Press, 2009.

\bibitem{dukhin_bulk_2009}
A.~S. Dukhin and P.~J. Goetz, ``Bulk viscosity and compressibility measurement
  using acoustic spectroscopy,'' {\em The Journal of Chemical Physics},
  vol.~130, p.~124519, Mar. 2009.

\bibitem{holmes_temperature_2011}
M.~J. Holmes, N.~G. Parker, and M.~J.~W. Povey, ``Temperature dependence of
  bulk viscosity in water using acoustic spectroscopy,'' {\em Journal of
  Physics: Conference Series}, vol.~269, p.~012011, Jan. 2011.

\bibitem{lin_bulk_2017}
J.~Lin, C.~Scalo, and L.~Hesselink, ``Bulk viscosity model for near-equilibrium
  acoustic wave attenuation,'' {\em arXiv:1707.05876 [physics]}, July 2017.

\bibitem{olver_nist_2010}
F.~W.~J. Olver and N.~I. of~Standards {and} Technology~(U.S.), eds., {\em
  {NIST} handbook of mathematical functions}.
\newblock Cambridge ; New York: Cambridge University Press : NIST, 2010.
\newblock OCLC: ocn502037224.

\bibitem{bass_atmospheric_1990}
H.~E. Bass, L.~C. Sutherland, and A.~J. Zuckerwar, ``Atmospheric absorption of
  sound: {Update},'' {\em The Journal of the Acoustical Society of America},
  vol.~88, pp.~2019--2021, Oct. 1990.

\bibitem{ainslie_simplified_1998}
M.~A. Ainslie and J.~G. McColm, ``A simplified formula for viscous and chemical
  absorption in sea water,'' {\em The Journal of the Acoustical Society of
  America}, vol.~103, pp.~1671--1672, Mar. 1998.
\newblock Publisher: Acoustical Society of America.

\bibitem{duck_physical_1990}
F.~A. Duck, {\em Physical properties of tissue: a comprehensive reference
  book}.
\newblock London: Academic Press, 1990.

\bibitem{bialek_biophysics:_2012}
W.~S. Bialek, {\em Biophysics: searching for principles}.
\newblock Princeton, NJ: Princeton University Press, 2012.

\end{thebibliography}
\bibliographystyle{ieeetr}

\appendix

\renewcommand{\thesubsection}{\Alph{section}.\arabic{subsection}}
\makeatletter
\renewcommand{\p@subsection}{}
\renewcommand{\p@subsubsection}{}
\makeatother

\definecolor{sTitleColorDark}{RGB}{40, 42, 46}
\definecolor{sLightBlue}{RGB}{129, 162, 190}

\newcommand{\scolorbox}[3]{
  \begin{tcolorbox}[
    colback=#1!10,
    colframe=#1,
    coltitle=sTitleColorDark,
    fonttitle=\sffamily\bfseries,,
    title=#2
  ]
  {#3}
  \end{tcolorbox}
}
\newcommand{\statement}[2][Note]{\scolorbox{sLightBlue}{#1}{#2}}

\section{Guide to the appendices}

The following appendix sections provide complete derivations of the energetic cost of communication (equation \ref{eq:cost_per_bit}) for each of the systems discussed in the main text.
Appendix \ref{app:sec:electrical} gives a detailed derivation of communication using electrical signaling between ion channels in membranes.
Appendix \ref{app:sec:diffusion} gives a detailed derivation of communication using diffusive signaling in 2D and 3D.
Appendix \ref{app:sec:acoustic} gives a detailed derivation of acoustic communication.

Computing the cost per bit involves first computing the dissipation kernel $\dissfn(\om)$, the noise in the output signal $S^{O}(\om)$, and the gain of the signaling network $|\ch^{OI}(\om)|^2$ for each of the system.
To remove redundancy in these calculations, three useful lemmas have been provided in appendix \ref{app:sec:lemmas}, which are reused in all of these derivations.

\subsection{Conventions}
Throughout all the appendices (as well as the main text), the following conventions are used:
\begin{itemize}
  \item Senders are denoted by $I$ and always located at the origin. Receivers are denoted by $O$ and centered at point $\bb r$
  \item $\ell=\ell(\om)$ refers to the characteristic lengthscale, defined differently for each system.
  \item $x$, $y$, and $z$ denote dimensionless lengthscale ratios. $x=\sgI/\ell$ is the rescaled sender size. $y=\sgO/\ell$ is the rescaled receiver size. $z=r/\ell$ is the rescaled transmission distance.
  \item Universal integral functions are subscripted by $U_D(x)$, $U_N(y)$, $U_S(z)$ when they correspond to the signal, noise, and dissipation respectively. Note the arguments of $x$, $y$, $z$: dissipation depends on $\sgI$, noise on $\sgO$, and signal on $r$
  \item Fourier transforms are implicit, so $\lm(k,\om)$ is assumed to be the space and time Fourier transform of $\lm(x, t)$.
    We use the Fourier convention: $f(\om) = \int\d t\, e^{-i\om t}f(t)$, $f(t) = \int\f{\d\om}{2\pi}e^{i\om t}f(\om)$.
\end{itemize}

\section{Electrical Signaling on a Membrane}
\label{app:sec:electrical}
Here we make explicit the derivation outlined in the main text.
This essentially requires two main steps.
The first is setting up the problem and solving for the dynamics of the system.
The second is analyzing the universal functions that the analysis produces.

\subsection{Setup}
As depicted in fig.~\ref{fig:fig1}, we suppose we have a 2D membrane which behaves like a capacitor with capacitance per unit area $c$ embedded in a bulk solution with conductivity $\al$. The surface charge density on the membrane is given by $\lm(x,y)$, defined such that $\<\lm\>=0$ (which simplifies the linear response calculation). No free charges are allowed to accumulate in the bulk.

The sender $I$, located at the origin, is the ion channel shown in blue which opens and closes, allowing the membrane to depolarize. This induces the input signal $I(t)$, the current flowing through the origin. The localized behavior of the sender is characterized by a current density $J(x,y,t)$ which disperses the current $I(t)$ in a Gaussian disk of radius $\sgI$:
\begin{align}
  \label{eq:es:input_signal}
  \bb J(x,y,t) = \f{e^{-\bb x^2/2\sgI^2}}{2\pi\sgI^2}I(t)\bb {\hat{z}}
  &&
  \tn{(A/m$^2$)}
\end{align}
The receiver $O$, located at $\bb r$ is the ion channel shown in yellow which detects the excess charge accumulated in its vicinity. Its localized detection is captured by a Gaussian disk of radius $\sgO$:
\begin{align}
  O(t) = \int\d x\d y\, e^{-(\bb x - \bb r)^2/2\sgO^2}\lm(x,y,t)
  &&
  \tn{(C)}
\end{align}

\subsection{Parameters}
\label{app:sec:el:paramters}
This system is described by a capacitance per area $c$ and a conductance per unit length $\al$.
In analogy with a traditional RC circuit, which has timescale $\tau=RC$, we instead have here an RC {\it velocity} given by:
\begin{equation}
  \nu_{RC} = \f{\al}{c} \approx 10^8\,\tn{$\mu$m/s}
\end{equation}
where the values of $\al$ and $c$ are taken to be $\al=10^{-2}$ $\Om^{-1}\tn{cm}^{-1}$~\cite{wang_role_2013}, $c=10^{-6}$ F/cm$^2$~\cite{gentet_direct_2000}.
This means that at frequency $\om$, the system has the characteristic lengthscale:
\begin{equation}
  \label{eq:es:lengthscale}
  \ell(\om) = \f{\al}{\om c}
\end{equation}
Which we will show determines the limit of how far signals at frequency $\om$ may be sent.
The logic of this limitation is very similar to the sense in which the RC timescale limits the admissible frequencies which may pass unimpeded through an RC circuit.
For reference, for the frequency range $f\in\{10^-3, 10^4\}$ Hz used in fig 3 from the main text, this characteristic lengthscale is in the range:
\begin{align*}
  f\in\{10^{-3}, 10^4\}\,\tn{Hz}
  &&
  \ell(2\pi f) \in\{1.6\,\tn{mm},\,16\,\tn{km}\}
\end{align*}
Thus, at sub $\tn{kHz}$ frequencies, this limiting lengthscale is quite large ($>5$ mm).

\subsection{Dynamics}

To obtain the full dynamics of this system, we need two response functions.
First we need $\ch^{\lm h}$, the susceptibility of the surface charge density which we can use with the fluctuation dissipation theorem to find the equilibrium noise~\cite{tong_lectures_2012}.
The second is $\ch^{\lm I}$, the linear response function of the surface charge density to the input signal~\cite{tong_lectures_2012}.

Following the usual prescription in linear response theory (see~\cite{tong_lectures_2012,zwanzig_nonequilibrium_2001}), we write down the energy of the system and add a field $h$ coupled to $\lm$ to find the response function $\ch^{\lm h}$. In this case the energy is that of a capacitor:
\begin{equation}
  H = \int\d x\d y\,\l[\f{1}{2}\f{\lm^2(\bb x)}{c} - h(\bb x)\lm(\bb x)\r]
\end{equation}
where the usual $H=Q^2/2C$ can be expressed in differential form under the assumption that the membrane width is much smaller than variations in $\lm$. Then the effective potential on the membrane is given by the functional derivative:
\[
  \ph(x,y,z=0^+) \equiv \f{\dl H}{\dl\lm} = \f{\lm(\bb x)}{c} - h(\bb x)
\]
In the bulk, the potential obeys Laplace's equation, which we calculate in $xy$-momentum space:
\[
  \nabla^2\ph(x,y,z) = 0
  \quad\to\quad
  \l[
    -k_x^2 + k_y^2 + \f{\d}{\d z}^2
  \r]\ph(\bb k,z) = 0
\]
Yielding solution:
\[
  \ph(\bb k, z\neq 0) = \ph(\bb k) e^{-k z}
  \qquad\quad
  k = \sqrt{k_x^2+k_y^2}
\]
Invoking continuity at $z=0$, we have:
\[
  \label{eq:es:potential_1}
  \ph(\bb k, z) =
  \l[
    \f{\lm(\bb k)}{c} - h(\bb k)
  \r]
  e^{-k z}
\]
In the absence of a signal, the current in the bulk in given by the differential Ohm's equation $\bb J_0(\bb x, z\neq 0) = -\al\div\ph$. The current at $z=0$ is zero since the membrane acts like a perfect insulator. Thus we can write the current using the Heaviside step function:
\[
  \bb J_0(\bb x, z) = -\al\nabla\ph(\bb x, z\neq 0)\Theta(z)
\]
We plug this into the continuity equation on the 3D charge density $\p_t\rh(x,y,z)=-\bb\nabla\cdot\bb J_0$:
\[
  \p_t \rh
  =
  \al \div^2\ph(\bb x,z\neq 0) \Theta(z)
  +
  \al \l[\div\ph(\bb x,z\neq 0)\cdot \dl(z)\hat{z}\r]
\]
By construction, $\div^2\ph(\bb x,z\neq 0)=0$ because it must satisfies Laplace's equation.
We then plug in $\rh(x,y,z) = \lm(\bb x)\dl(z)$ and drop the delta functions:
\[
  \p_t \lm(x,y)
  =
  \al \f{\d}{\d z}\ph(\bb x,z\neq 0)
\]
Moving to Fourier space and plugging in eq. (\ref{eq:es:potential_1}) we are left with:
\[
  \p_t\lm(\bb k) = -\f{\al}{c} k\lm(\bb k) + \al k h(\bb k)
\]
Which is the basic dynamics in the absence of an input signal. When a weak input signal $J(\bb k,t)$ of the form given in eq. (\ref{eq:es:input_signal}) is present, we have:
\[
  \p_t\lm(\bb k)
  =
  -\f{\al}{c} k\lm(\bb k)
  + \al k h(\bb k)
  +
  J(\bb k,t)
\]
which is exactly eq. (3) from the main text.
Finally, we move to frequency space to obtain the full dynamics:
\begin{equation}
  \l[i\om + \f{\al k}{c}\r]\lm(\bb k,\om) = \al k h(\bb k,\om) + J(\bb k,\om)
\end{equation}
From these dynamics we can extract the susceptibility $\ch^{\lm h}$ and input response $\ch^{\lm I}$, defined to satisfy $\<\lm(\bb k,\om)\>=\ch^{\lm h}(\bb k,\om)h(\bb k,\om)$ and $\<\lm(\bb k,\om)\>=\ch^{\lm I}(\bb k,\om)I(\om)$
\begin{align}
  \label{eq:es:response_functions}
  \ch^{\lm h}(\bb k,\om) = \f{ck}{k + i\ell^{-1}}
  &&
  \ch^{\lm I}(\bb k,\om) = \f{c}{\al}\f{e^{-k^2\sgI^2/2}}{k + i \ell^{-1}}
\end{align}
and the related response function of $\lm$ to $J$:
\[
  \ch^{\lm J}(\bb k,\om) = \f{c}{\al}\f{1}{k + i \ell^{-1}}
\]
It's worth noting that this is where the dependence on the characteristic lengthscale $\ell=\al/c\om$ of the system comes from.

\subsection{Signal}
Here the goal is to compute $|\ch^{OI}(\om)|^2$, the gain of the signaling network.
We start with the response function of $\lm$ to $J$: $\ch^{\lm J}(\bb k,\om) = \f{c}{\al}\f{1}{k + i\ell^{-1}}$
Invoking eq. \ref{eq:lemma:output_response_function_2d}, we find:
\begin{equation*}
    \ch^{OI}(\om)
    =
    \f{c}{\al}\f{\sgO^2}{r}
    \int_0^\infty\d u\,
    \f{u J_0(u)}{u + i r/\ell}
\end{equation*}
The gain function is this squared, which we can write in terms of the scaling portion and a universal integral function:
\begin{equation}
  \label{eq:es:signal_raw}
  \begin{split}
  |\ch^{OI}(\om)|^2
  &=
  \f{c^2\sgO^4}{\al^2 r^2}
  \l|U^\el_S\fp{r}{\ell(\om)}\r|^2
  \\
  U^\el_S(z)
  &\equiv
  \int_0^\infty\d u\,
  \f{u J_0(u)}{u + i z}
  \end{split}
\end{equation}
The important fact here is that when $r\ll \ell(\om)$, we have $U^\el_S(r/\ell)=1$ and can be completely ignored.
We can thus think of $U^\el(r/\ell)$ as a small correction that only becomes relevant once the transmission distance approaches the characteristic lengthscale.
This can be demonstrated by breaking $U^\el_S$ into its real and imaginary components, $\l|U^\el_S(z)\r|^2 = U^\el_{S,\tn{Re}}(z)^2 + U^\el_{S,\tn{Im}}(z)^2$, which each have analytic solutions:
\begin{equation}
  \label{eq:el:signal-function-exact}
  \begin{split}
  U^\el_{S,\tn{Re}}(z)
  &\equiv
    \int_0^\infty\d u\,
    \f{u^2 J_0(u)}{u^2 + z^2}
    =
    1 + \f{\pi z}{2}M_0(z)
  \\
  U^\el_{S,\tn{Im}}(z)
  &\equiv
    \int_0^\infty\d u\,
    \f{uz J_0(u)}{u^2 + z^2}
    = z K_0(z)
  \end{split}
\end{equation}
where $M_0$ is a modified Struve function (see 11.2.6 from~\cite{olver_nist_2010}) and $K_0$ is a modified Bessel function.
We can get the $z\to 0$ behavior of $U^\el_{S,\tn{Re}}$ using a power series expansion of $M_0(z)$ (see 11.2.6, 11.2.2, 10.25.2 from~\cite{olver_nist_2010}),
and the behavior of $U^\el_{S,\tn{Im}}$ using the asymptotic expression for $K_0$ (see 10.30.3 from~\cite{olver_nist_2010}):
\begin{align*}
  U^\el_{S,\tn{Re}}(z)
  &=
  1 + z\sum_{k=0}^\infty\l[
    \f{z^{2k+1}}{(2k+1)!!^2}
    -
    \f{\pi}{2}\f{z^{2k}}{2^{2k}k!^2}
  \r]
  \sim
  1 - \f{\pi z}{2}
  \\
  U^\el_{S,\tn{Im}}(z)
  &=
  zK_0(z)
  \sim
  -z\log z
\end{align*}
and therefore the approximation $U^\el_S(z\ll 1)=1$ is well-justified.
We can also obtain the $z\to\infty$ behavior using the asymptotic expressions for $M_0$ (see 11.6.2 from~\cite{olver_nist_2010}) and $K_0$ (10.25.3 from~\cite{olver_nist_2010}):
\begin{align*}
  U^\el_{S,\tn{Re}}(z\to\infty)
  &\sim
  -\sum_{k=1}^\infty (-1)^{k}\f{\Gamma(1/2+k)}{\Gamma(1/2-k)}\f{2^{2k}}{z^{2k}}
  \sim -\f{1}{z^2}
  \\
  U^\el_{S,\tn{Im}}(z\to\infty)
  &\sim
  \sqrt{\f{\pi z}{2}}e^{-z}
\end{align*}
Plugging these into $|U^\el_S|^2=|U^\el_{S,\tn{Re}}|^2+|U^\el_{S,\tn{Im}}|^2$, we obtain the $z\to 0$ and $z\to\infty$ asymptotic behavior of $|U^\el_S|^2$:
\begin{align}
  \label{eq:electrical:signal-app-1}
  \l|U^\el_S(z\to 0)\r|^2
    &\sim
    1 - \pi z +\OO(z^2\log(z)^2)
  \\
  \label{eq:electrical:signal-app-2}
  \l|U^\el_S(z\to\infty)\r|^2
    &\sim \f{1}{z^4}
\end{align}
In producing the plots in the main paper, this function is evaluated numerically.
We only derived these asymptotic limits to give the scaling behavior in the $z\ll 1$ and $z\gg 1$ regimes.

\subsection{Noise}
Here we compute $S^O(\om)$, the fluctuations in the output signal in the absence of an input arising from thermal noise. The fluctuation dissipation theorem~\cite{tong_lectures_2012} tells us that since $h$ is conjugate to $\lm$ in the energy, the fluctuations in $\lm$ are given by:
\begin{equation*}
  S^\lm(\bb k,\om)
  = -\f{2\kbt}{\om}\Im{\ch^{\lm h}(\bb k,\om)}
  = \f{1}{\al}\f{2 c^2 \kbt k}{k^2 + (\om c/\al)^2}
\end{equation*}
We then invoke eq. (\ref{eq:lemma:output_power_spectrum_2d}) to get $S^O(\om)$:
\begin{equation*}
  S^O(\om)
  =
  \f{4\pi c^2 \sgO^3\kbt}{\al}
  \int_0^\infty\d u\,
    \f{u^2e^{-u^2}}{u^2 + (\sgO/\ell(\om))^2}
\end{equation*}
This integral has an analytic solution which permits us to write the noise in terms of a scaling portion and a universal function:
\begin{equation}
  \label{eq:es:noise_raw}
  \begin{split}
  S^O(\om) &= \kbt\f{2\pi^{3/2}c^2\sgO^3}{\al}U^\el_N\fp{\sgO}{\ell(\om)}
  \\
  U^\el_N(y)
  &\equiv
  1 - \sqrt{\pi} y e^{y^2}\erfc(y)
  \end{split}
\end{equation}
When $\sgO\ll \ell$, the noise correction function $U^\el_N\to 1$ and can be completely ignored.
This can be seen using the series expansion of $\erfc$:
\begin{align*}
  U^\el_N(y)
  &\equiv
  1 - \sqrt{\pi} y e^{y^2}\erfc(y)
  \\
  &=
  1 - \sqrt{\pi} y e^{y^2}\l[1 - \f{2}{\sqrt{\pi}}y + \OO(y^3)\r]
  \\
  &=
  1 - \sqrt{\pi} y + \OO(y^2)
\end{align*}
And thus when the receiver radius $\sgO$ is much smaller than the characteristic lengthscale $\ell$, we may ignore the contribution of $U^\el_N$.

\subsection{Dissipation}
The dissipation in this system is Ohmic: we run current density $J(\bb x,t)$ through the membrane, which has some electrical potential density difference $\lm(\bb x)/c$.
Like an ordinary circuit, this requires power $\int\d^2\bb x\, J(\bb x)\lm(\bb x)/c \sim IV$.
However, it is useful to analyze this system explicitly in terms of the capacitive energy since there is a symmetry between dissipation in electrical signaling and the other systems under consideration.
The energy stored in the capacitor is given by:
\[
  H = \int\d^2\bb x\, \f{1}{2c}\lm^2(\bb x)
\]
And thus the energetic dissipation required to run the network is given by:
\[
  \<\tn{Diss}(t)\>
  = \l\<\f{\p H}{\p\lm}\fp{\p\lm}{\p t}_{J}\r\>
  = \int\f{\d^2\bb x}{c}\<\lm(\bb x,t)J(\bb x,t)\>
\]
We then invoke eq. (\ref{eq:lemma:dissipation_kernel}) with prefactor $\CC=1/c$, giving us:
\begin{align*}
  \dissfn(\om) &= \f{1}{(2\pi)^3 \al \sgI}
                 \int\d^2\bb u\,
                 \f{u e^{-u^2}}{u^2 + (\sgI\om c/\al)^2}
\end{align*}
Which we can write in terms of a scaling portion and a universal integral function:
\begin{equation}
  \label{eq:es:diss_raw}
  \begin{split}
  \dissfn(\om) &= \f{1}{8\pi^{3/2}\al\sgI}U^\el_D\fp{\sgI}{\ell(\om)}
  \\
  U^\el_D(x)
  &=
  U^\el_N(x)
  \equiv
    \f{2}{\sqrt{\pi}}
    \int_0^\infty\d u\,
    \f{u^2e^{-u^2}}{u^2 + x^2}
  \end{split}
\end{equation}
where $U^\el_D$ is actually the same universal integral function as $U^\el_N$, the integral function found in the noise term.
Likewise, we have that $U^\el_D(\sgI/\ell) = 1$ when $\sgI\ll \ell$.

\subsection{Energy Cost Per Bit}
\label{app:sec:es:energy-cost}
Here we combine our results for the electrical signal (eq. (\ref{eq:es:signal_raw})), the noise (eq. (\ref{eq:es:noise_raw})), and the dissipation (eq. (\ref{eq:es:diss_raw})) into our expression for the energetic cost per bit (eq. (\ref{eq:cost_per_bit})). We find that the energetic cost per bit to send information between ion channels in a 2D membrane at frequency $\om$ in $\kbt$/bit is given by:
\begin{align}
  \label{eq:es:cost_raw}
  \costfn^\el &=
    \pi\log 2\,
    \f{r^2}{\sgI\sgO}
    \f{U^\el_D\fp{\sgI}{\ell(\om)}U^\el_N\fp{\sgO}{\ell(\om)}}{\l|U_S^\el\fp{r}{\ell(\om)}\r|^2}
\end{align}
where $r$ is the transmission distance, $\sgI$ is the sender radius, $\sgO$ is the receiver radius, and $\ell(\om)=\al/c\om$ is the characteristic lengthscale.
The functions $U^\el_D$, $U^\el_N$, and $U^\el_S$ are the integral functions derived earlier which are all equal to $1$ when their arguments are smaller than 1.

When the signaling lengthscales are smaller than the characteristic lengthscale (i.e. $\sgO$, $\sgI$, $r$ $\ll \ell(\om)$), the universal functions can all be ignored since $U^\el_D=U^\el_N=U^\el_S = 1$.
This condition can be reduced to requiring that $r\ll\ell(\om)$ since the concept of signaling is not well defined when the sender/receiver size are comparable or larger than the transmission distance (i.e. we always have $\sgI$, $\sgO$ $\ll r$).
This yields the expression for the energetic cost below the characteristic lengthscale:
\begin{align}
  \label{eq:es:cost_scaling}
  \tn{when $r$ $\ll$ $\ell(\om)$}
  &&
  \costfn^\el = \pi\log 2\,\f{r^2}{\sgI\sgO}
  \qquad\tn{($\kbt$/bit)}
\end{align}
A surprisingly simple result given the complexity of its ingredients.
Note that it does not depend on the physical constants $\al$ or $c$.
These constants only matter in determining the limiting behavior via the characteristic lengthscale.

Above the characteristic lengthscale, we need to handle the integral functions.
In this system as well as all the signaling systems, we need to use a mix of analytic, numerical, and asymptotic approaches to evaluate these.
We can view these integral functions as corrections to the energetic cost that reflect limiting physical behavior of the system near the characteristic lengthscale.

In each of the universal functions, $U^\el_S(r/\ell)$, $U^\el_N(\sgO/\ell)$, $U^\el_D(\sgI/\ell)$, the value only deviates from $1$ once its argument begins to approach $1$. In a well-defined signaling process, the sender/receiver sizes should be significantly smaller than the transmission distance ($\sgI$, $\sgO$ $\ll$ $r$).
Thus the argument to $U^\el_S$ is always much larger than for the other two.
This means the first order corrections to eq. (\ref{eq:es:cost_scaling}) are dominated by contributions from the universal function for the signal: $U^\el_S$, which has the analytic expression given by eq.~\ref{eq:el:signal-function-exact}.
A plot of the real and imaginary components of $U^\el_S$ is shown below

\begin{figure}[H]
\includegraphics[width=0.8\textwidth]{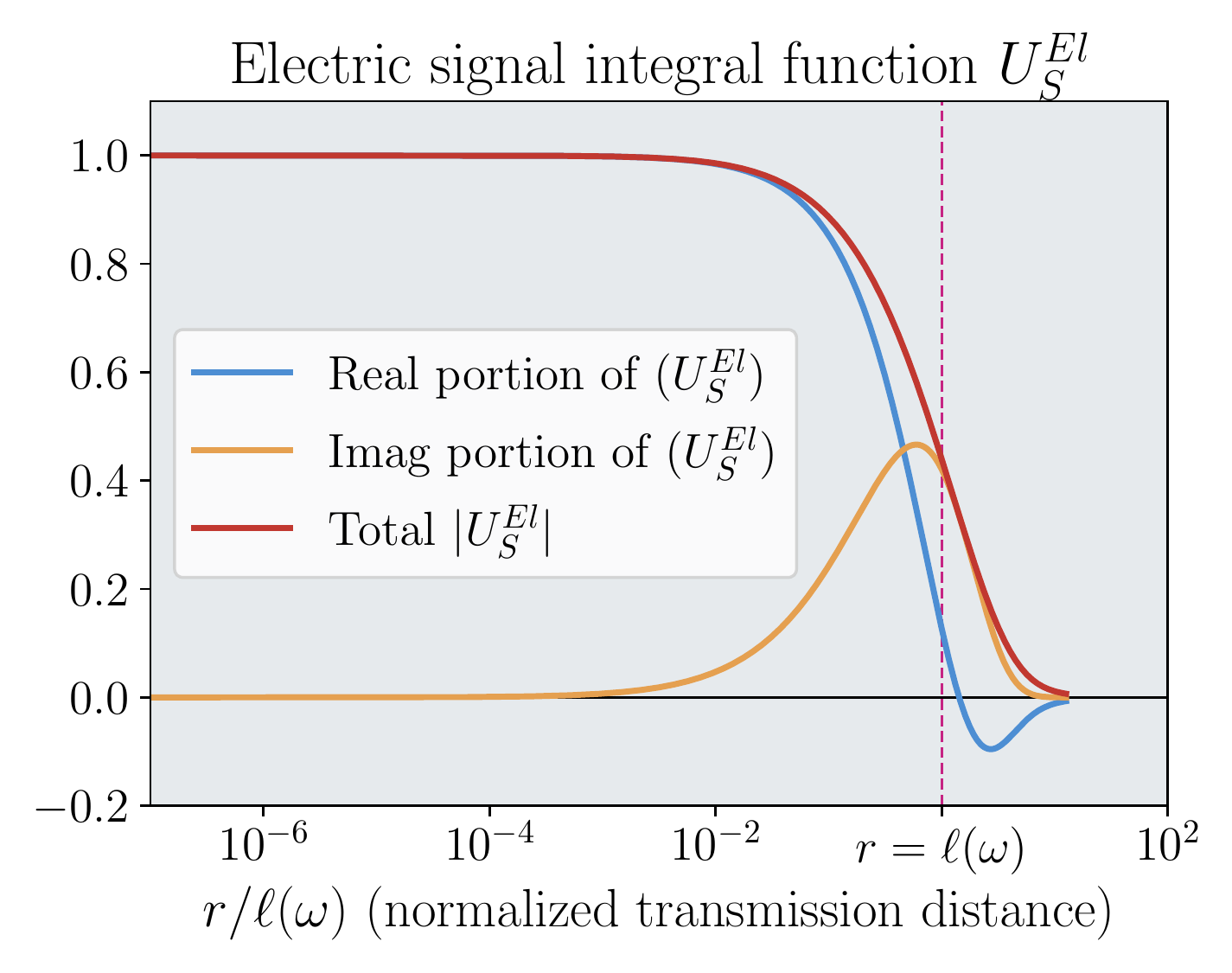}
\caption{
    The integral function contribution to the electrical signal transmission coefficient. This plots $U^\el_S$ as a function of the transmission distance normalized by the characteristic lengthscale. In blue is the real component, in yellow is the imaginary component, and in red is the total value
}\label{fig:es:usignal}
\end{figure}

From fig. \ref{fig:es:usignal}, we can develop a clearer picture of the underlying physics. The imaginary portion of $U^\el_S$ indicates the phase difference between the sender and receiver.
Well below the characteristic lengthscale, the entire system is in phase and there is no loss in signal strength.
As the transmission distance approaches the characteristic lengthscale, the signal becomes highly attenuated due to the collapse in resolution between subsequent phases.
This is seen by the rapid falloff in the real portion of $U^\el_S$, which is the dominant contribution.
Just beyond the characteristic lengthscale, the real portion of $U^\el_S$ becomes negative, but non-zero indicating that the communication is out of phase and highly attenuated.
Even further in the $r > \ell(\om)$ regime, the subsequent phases of the signal merge during transmission and the signal becomes fully attenuated.

As mentioned, the integral functions for the noise and dissipation are less relevant in evaluating eq. (\ref{eq:es:cost_raw}), the energetic cost per bit, but they are included in the calculations which produced the plots shown in the main text.

\section{Diffusive Signaling in 2D and 3D}
\label{app:sec:diffusion}

Here we do an explicit analysis of the energetic cost per bit to send information via diffusion in 2D and 3D.
We proceed in a similar structure as with the electrical analysis.

One caveat to keep in mind is that this setup assumes the sender injects/removes particles.
This is slightly different from more realistic modification-diffusive communication system where the sender consumes energy to modify/de-modify the state of the carrier molecules.
The system described here turns out to be the maximally efficient limit of modification-diffusive communication, which is briefly discussed in section \ref{app:sec:diffusion:modify-diffuse}.

\subsection{Setup}
As depicted in fig.~\ref{fig:fig1}d, we suppose we have a 2D membrane or 3D bulk with a particle density $\lm(\bb x,t)$, defined so that $\<\lm\>=0$.
The equilibrium dynamics are entirely produced by diffusion with some diffusion constant $D$.

The input and output signals are defined similarly as with electrical signaling.
We let $d=2$, $3$ be the dimensionality of the system.
The sender, depicted in blue, injects/removes particles at the origin with rate $I(t)$, localized in a Gaussian disk/sphere of radius $\sgI$:
\begin{align}
  \label{eq:df:input_signal}
  J = \f{e^{-\bb x^2/2\sgI^2}}{(2\pi\sgI^2)^{d/2}}I(t)
  &&
  \tn{(particles/s/m$^d$)}
\end{align}
The receiver, shown in yellow, is a disk/sphere located at some position $\bb r$.
The output signal is the number of excess particles localized in a Gaussian disk/sphere of radius $\sgO$ around the receiver:
\begin{align}
  \label{eq:df:output_signal}
  O(t) = \int\d^d\bb x \, e^{-(\bb x - \bb r)^2/2\sgO^2}\lm(\bb x,t)
  &&
  \tn{(particles)}
\end{align}

\subsection{Parameters}
This system is described by $D$, the diffusion constant.
In time $t$, the typical distance traveled by a particle diffusing in $d$ dimensions is given by: $\<\Delta x^2\> = 2dDt$.
Thus, for a signal sent at frequency $\om$, particles may travel a distance:
\[
  \sqrt{\<\Delta x^2\>} = \sqrt{2dD\f{T}{2}} = \sqrt{2\pi d D/\om}
\]
in one phase (i.e. half-period) of the signal.
This means that the characteristic lengthscale of diffusive signaling at frequency $\om$ is given by:
\begin{equation}
  \label{eq:diffusion:lengthscale}
  \ell(\om) = \sqrt{D/\om}
\end{equation}
Signals sent further than this lengthscale are strongly attenuated because of phase diffusion.
This fact will explicitly arise from the calculation of $\ch^{OI}$, the gain of the diffusive signaling process, in section \ref{app:sec:diffusion:signal}.

\subsection{Dynamics}
The dynamics of this system can be analytically solved in a straightforward manner.
One can write down the particle density in terms of each particle:
\[
  \lm(\bb x,t) = \sum_{i} \dl(\bb x-\bb x_i(t))
\]
and then use diffusive correlation functions to generate the response function and equilibrium fluctuations.
However, to keep a symmetry between the various systems under consideration, we take an alternative approach that uses the energy of the system and a chemical potential.

We start by denoting $\rh(\bb x)$ to be the true particle density such that $\lm(\bb x) = \rh(\bb x) - \rh_0$, where $\rh_0$ is the average density in the absence of a signal (everything done here is done in the absence of an input signal which is introduced later).
The entropy of the system is given by:
\begin{align*}
  -\bt S &= \int\d^d\bb x\,\rh\log\rh
     = \int\d^d\bb x\,(\lm+\rh_0)\log(\lm+\rh_0)
     \\
     &= \int\d^d\bb x\,\l[(\lm+\rh_0)\log\l(1 + \f{\lm}{\rh_0}\r) + (\lm+\rh_0)\log\rh_0\r]
\end{align*}
Next we note that $\int\lm\log\rh_0 = 0$ and that $\int\rh_0\log\rh_0$ is just the constant zero-point entropy of the system.
This allows us to write down the free energy density of the system (relative to the zero-point free energy):
\begin{align*}
  \bt \f{\d F}{\d V}
  &\equiv
    -\bt\f{\d S}{\d V} - \bt\lm(\bb x)h(\bb x)
  \\
  &=
      (\lm(\bb x)+\rh_0)\log\l(1 + \f{\lm(\bb x)}{\rh_0}\r)
      -
      \bt\lm(\bb x)h(\bb x)
\end{align*}
where $h(\bb x)$ is an artificial external field which couples to $\lm(\bb x)$, analogous to $h$ in the electrical derivation, introduced solely to compute the susceptibility of the particle density.
The chemical potential $\mu$ can be obtained via the functional derivative of $F$ with respect to $\lm$:
\[
  \bt\mu \equiv \f{\dl (\bt F)}{\dl \lm}
  =
  1 + \log\l(1 + \f{\lm(\bb x)}{\rh_0}\r) - \bt h(\bb x)
\]
This chemical potential induces a current density $J_0$
\begin{align*}
  J_0 = -\bt D\rh\div\mu
  = -D\l[\div\lm - \bt \rh\div h\r]
\end{align*}
which obeys a conservation equation $\p_t\lm = -\div J$:
\begin{align*}
  \p_t\lm
  =
  D\div^2\lm - D\bt(\div\lm)(\div h) - D\bt(\rh_0+\lm)\div^2 h
\end{align*}
Now we also introduce the (non-conservative) input signal $J(t)$:
\begin{equation}
  \label{eq:diffusion:full-dynamics}
  \begin{split}
  \p_t\lm
  =
  &+D\div^2\l(\lm - \bt\rh_0 h\r)
  \\
  &- D\bt\l[(\div\lm)(\div h) + \lm\div^2 h\r]
  + J(\bb x, t)
  \end{split}
\end{equation}
Which gives the complete real-space dynamics of this system.
We then Fourier transform to frequency/momentum space:
\begin{align*}
  i\om\lm(\bb k,\om)
  =
  &-Dk^2\l[
    \lm(\bb k,\om)
    -
    \bt \rh_0 h(\bb k,\om)
  \r]
  + J(\bb k,\om)
  \\
  &-
  D\bt\FF[(\div\lm)(\div h) + \lm\div^2 h](\bb k,\om)
\end{align*}
Reading off the linear terms, we can extract the susceptibility $\ch^{\lm h}$ and input response $\ch^{\lm I}$~\cite{tong_lectures_2012}, defined by $\<\lm(\bb k,\om)\>=\ch^{\lm h}(\bb k,\om)h(\bb k,\om)$ and $\<\lm(\bb k,\om)\>=\ch^{\lm I}(\bb k,\om)I(\om)$:
\begin{align}
  \label{eq:df:response_functions}
  \ch^{\lm h}(\bb k,\om) = \f{\bt\rh_0k^2}{k^2 + i\ell^{-2}}
  &&
  \ch^{\lm I}(\bb k,\om) = \f{1}{D}\f{e^{-k^2\sgI^2/2}}{k^2 + i\ell^{-2}}
\end{align}
where we have also exploited the Fourier transform of $J(\bb x,t)$ (eq. (\ref{eq:df:input_signal})) and used the characteristic lengthscale $\ell=\sqrt{D/\om}$.

\subsection{Signal}
\label{app:sec:diffusion:signal}
Here the goal is to compute $|\ch^{OI}(\om)|^2$, the gain of the signaling network.
We start with the response function of $\lm$ to $J$: $\ch^{\lm J}(\bb k,\om) = \f{1}{D}\f{1}{k^2 + i\om/D}$.
\\

\noindent
{\bf 2D Result}
Using eq. (\ref{eq:lemma:output_response_function_2d}), we can show that in 2D, $\ch^{OI}(\om)$ is given by:
\begin{align*}
  \ch^{OI}(\om)
  =
    \f{\sgO^2}{D}
    \int_0^\infty\d u\,
    \f{uJ_0(u)}{u^2 +i(r/\ell)^2}
\end{align*}
Which allows us to write $|\ch^{OI}(\om)|^2$ in terms of a scaling term and an integral function which has a simple analytic form:
\begin{align}
  \label{eq:dff:signal_raw}
  \tn{(2D)}
  &&
  \begin{split}
  |\ch^{OI}(\om)|^2
  &=
    \f{\sgO^4}{D^2}
    \l|U^\dff_S\fp{r}{\ell}\r|^2
    \\
  U^\dff_S(z)
    &\equiv
    K_0(\sqrt{i}z)
  \end{split}
\end{align}
where $K_0$ is a Bessel function.
We note that unlike electrical signaling, 3D diffusion, and acoustic signaling, this function diverges for small arguments.
Therefore, we cannot ignore $U^\dff_S(r/\ell)$ in the $r \ll \ell$ limit.
Thus, in order to obtain the leading behavior of the energetic cost, we have to look at the asymptotic behavior of this function.
For $z\to 0$, $K_0(\sqrt{i}z)$ has the asymptotic form given by 10.30.3 from~\cite{olver_nist_2010}:
\begin{align}
  \label{eq:dff:signal_app}
  \tn{for $z\to 0$}
  &&
  \begin{split}
  U^\dff_S(z) &\sim
    -\ln(z) - i\f{\pi}{4}
  \\
  \l|U^\dff_S\r|^2 &\sim \ln(z)^2 + \fp{\pi}{4}^2
  \end{split}
\end{align}
For $z\to \infty$, $K_0(\sqrt{i}z)$ has the asymptotic form given by 10.25.3 from~\cite{olver_nist_2010}:
\begin{align}
  \label{eq:dff:signal_app_2}
  \tn{for $z\to\infty$}
  &&
  \begin{split}
  U^\dff_S(z) &\sim \sqrt{\f{\pi}{2\sqrt{i}z}}e^{-\sqrt{i}z}
  \\
  \l|U^\dff_S\r|^2 &\sim \f{\pi}{2z}e^{-\sqrt{2}z}
  \end{split}
\end{align}
The first relation gives us the scaling form of $U^\dff_S$ for transmission distances below the characteristic lengthscale $\ell$.
The second relation gives the $e^{-\sqrt{2}z}$ exponential attenuation incurred for distances above $\ell$, which is found in all of the studied systems.
\\

\noindent
{\bf 3D Result}
Using eq. (\ref{eq:lemma:output_response_function_3d}), the output response function in 3D is given by:
\begin{align*}
    \ch^{OI}(\om) = \sqrt{\f{2}{\pi}}\f{\sgO^3}{Dr}
      \int_0^\infty\d u\f{u\sin u}{u^2+i(r/\ell)^2}
\end{align*}
This can be solved analytically using contour integration by rewriting $2i\sin u=e^{i u}-e^{-iu}$ and then closing the two terms over the upper and lower semi-circles.
This allows us to write $|\ch^{OI}(\om)|^2$ in terms of a scaling term and a correction function:
\begin{align}
  \label{eq:dfff:signal_raw}
  \tn{(3D)}
  &&
  \begin{split}
  |\ch^{OI}(\om)|^2
  &=
  \f{\pi}{2}\f{\sgO^6}{D^2 r^2}\l|U^\dfff_S\fp{r}{\ell}\r|^2
  \\
  \l|U^\dfff_S(z)\r|^2
    &\equiv e^{-\sqrt{2}z}
  \end{split}
\end{align}
where we can ignore the contribution from $U^\dfff_S$ when the transmission distance is smaller than the characteristic lengthscale ($z\ll 1$).

\subsection{Noise}
\label{app:sec:diffusion:noise}
Here we compute $S^O(\om)$, the fluctuations in the output signal in the absence of an input arising from thermal noise. Just as with electrical signaling, the fluctuation dissipation theorem~\cite{tong_lectures_2012} tells us that since $h$ is conjugate to $\lm$ in the energy, the fluctuations in $\lm$ are given by:
\begin{equation*}
  S^\lm(\bb k,\om)
  = -\f{2\kbt}{\om}\Im{\ch^{\lm h}(\bb k,\om)}
  = \f{2\rh_0}{D}\f{k^2}{k^4 + \ell^{-4}}
\end{equation*}

\noindent
{\bf 2D Result}
We then invoke eq. (\ref{eq:lemma:output_power_spectrum_2d}) to get $S^O(\om)$ in 2D:
\begin{align*}
  S^O(\om)
  &=
  \f{4\pi\rh_0\sgO^4}{D}\int_0^\infty\d u\,\f{u^3 e^{-u^2}}{u^4 + (\sgO/\ell(\om))^4}
\end{align*}
which we can write in terms of the scaling portion and a universal integral function:
\begin{align}
  \label{eq:dff:noise_raw}
  \tn{(2D)}
  &&
  \begin{split}
  S^O(\om)
  &=
  \f{2\pi\rh_0\sgO^4}{D} U^\dff_N\fp{\sgO}{\ell(\om)}
  \\
  U^\dff_N(y)
  &\equiv
  2\int_0^\infty\d u\f{u^3e^{-u^2}}{u^4+y^4}
  \end{split}
\end{align}
Just like the 2D diffusive noise, this function also diverges as $y\to 0$,
and thus the leading behavior is obtained by looking at its asymptotic expansion.
The noise correction function has an analytic solution which we can write in terms of the auxiliary function $g$ defined in 6.2.18 from ref.~\cite{olver_nist_2010}:
\begin{align*}
  U^\dff_N(y)
  = g(y^2)
  = \f{\pi}{2}\sin(y^2) - \cos(y^2)\tn{Ci}(y^2)- \sin(y^2)\tn{Si}(y^2)
\end{align*}
where Ci/Si are the cosine and sine integral functions.
Then using the series expansions of Ci/Si given by 6.6.5-6 from~\cite{olver_nist_2010} and the asymptotic behavior of $g$ given by 6.12.4, we obtain the asymptotic behavior:
\begin{align}
  \label{eq:dff:noise_app}
  \tn{for $y\to0$}
  &&
  U^\dff_N(y) &\sim  - (\ga + \ln(y^2))
  \\
  \label{eq:dff:noise_app_2}
  \tn{for $y\to\infty$}
  &&
  U^\dff_N(y) &\sim  \f{1}{y^4}
\end{align}
where the first relation is relevant whenever the receiver size $\sgO$ is much smaller than the characteristic lengthscale $\ell$.
\\

\noindent
{\bf 3D Result}
In 3D, we use eq. (\ref{eq:lemma:output_power_spectrum}) to get $S^O(\om)$:
\begin{align*}
  S^O(\om)
  &=
  \f{8\pi\rh_0\sgO^5}{D}\int_0^\infty\d u
  \f{u^4e^{-u^2}}{u^4 + (\sgO/\ell(\om))^4}
\end{align*}
which we write in terms of the scaling portion and a universal function:
\begin{align}
  \label{eq:dfff:noise_raw}
  \tn{(3D)}
  &&
  \begin{split}
  S^O(\om)
  &=
  \f{4\pi^{3/2}\rh_0\sgO^5}{D}U^\dfff_N\fp{\sgO}{\ell(\om)}
  \\
  U^\dfff_N(y)
  &\equiv
  \f{2}{\sqrt{\pi}}
  \int_0^\infty\d u
  \f{u^4e^{-u^2}}{u^4 + y^4}
  \end{split}
\end{align}
Where the normalization of $U^\dfff_N$ is chosen such that $U^\dfff_N(y\ll 1)=1$ is a small correction that may be ignored when the receiver size is smaller than the characteristic lengthscale.
We can see this by looking at its analytic solution:
\begin{align*}
  U^\dfff_N(y) = 1 + \sqrt{2\pi} y
    &\l[\l(F_S\l(\sqrt{\f{2}{\pi}}y\r) - \f{1}{2}\r)\cos(y^2)\r.
    \\
    &-\l.
    \l(F_C\l(\sqrt{\f{2}{\pi}}y\r) - \f{1}{2}\r)\sin(y^2)\r]
\end{align*}
where $F_S$ and $F_C$ are the Fresnel sine and cosine integrals.
To find the $y\to 0$ behavior, we use the power series expansions of $F_C$ and $F_S$ (see 7.6.5 and 7.6.7 from~\cite{olver_nist_2010}) to find the power series for $U^\dfff_N$:
\begin{align*}
  U^\dfff_N(y)
  =
  1
  &+ \sqrt{\f{\pi}{2}}y\l[\sin(y^2)-\cos(y^2)\r]
  -
  4y^4\sum_{n=0}^\infty \f{(-1)^n (4y^4)^n}{(4n+3)!!}
\end{align*}
To find the $y\to\infty$ behavior, we recognize that $U^\dfff_N(y)$ can be written in terms of the auxiliary function $f$ defined in 7.2.10 from ref.~\cite{olver_nist_2010}:
\begin{align*}
  U^\dfff_N(y)
  =
  1 - \sqrt{2\pi} y f\l(\sqrt{\f{2}{\pi}}y\r)
\end{align*}
Then using the above power series as well as the asymptotic behavior of $f$ given by 7.12.2 from~\cite{olver_nist_2010}, we obtain the asymptotic limits:
\begin{equation}
  \label{eq:dfff:noise_scaling}
  \begin{split}
  U^\dfff_N(y\to 0)
  &\sim
  1 - \sqrt{\f{\pi}{2}}y\cos^2(y^2)
  \\
  U^\dfff_N(y\to\infty)
  &\sim
  \f{3}{4 y^4}
  \end{split}
\end{equation}
This justifies our earlier statement that $U^\dfff_N(y\to0)=1$.

\subsection{Dissipation}
\label{app:sec:diffusion:dissipation}
The dissipation in this system results from the entropic cost of creating particles at high concentration and degrading them at low concentration.
The dissipation is derived in much the same way as with electrical signaling.
The free energy of the system in the absence of an external field can be written:
\[
  F = \kbt\int\d^d\bb x\,(\rh_0+\lm)\log(\rh_0+\lm)
\]
Thus the change in free energy incurred by running the signal process is:
\begin{align*}
  \<\diss\>
  &= \l\<\f{\p F}{\p \lm}J(x,t)\r\>
  \\
  &= \kbt\l\<\int\d^d\bb x\,\l[\log\l(1+\f{\lm}{\rh_0}\r)+1+\log(\rh_0)\r]J(x,t)\r\>
  \\
  &= \kbt\int\d^d\bb x\l\<\log\l(1+\f{\lm}{\rh_0}\r)J(x,t)\r\>
\end{align*}
where we have used the fact that $\<J(t)\>=0$, i.e. the total particle count is conserved over long time periods.
Since we are in the weak signal regime, we expect the fluctuations in particle density to be small with respect to the average density.
We can therefore expand the log, yielding:
\[
  \<\diss\>
  \approx
  \f{\kbt}{\rh_0} \int\d^d\bb x\,
  \l\< \lm(\bb x,t) J(\bb x,t)\r\>
\]
We can now use eq. (\ref{eq:lemma:dissipation_kernel}) with prefactor $\CC=\kbt/\rh_0$ to get the dissipation kernel:
\[
  \dissfn(\om)
  =
  \f{\kbt\sgI^2}{(2\pi)^{d+1} \rh_0D\sgI^{d}}
  \int\d^d\bb u\f{u^2e^{-u^2}}{u^4+(\sgI/\ell)^4}
\]

\noindent
{\bf 2D Result}
In two dimensions the dissipation kernel becomes:
\[
  \dissfn(\om)
  =
  \f{\kbt}{4\pi^2 \rh_0D}
  \int_0^\infty\d u\f{u^3e^{-u^2}}{u^4+(\sgI/\ell)^4}
\]
which we write in terms of a scaling portion and a universal integral function, which is the same as the universal function for the noise in 2D:
\begin{align}
  \label{eq:dff:diss_raw}
  \tn{(2D)}
  &&
  \begin{split}
  \dissfn(\om)
  &=
  \f{\kbt}{8\pi^2\rh_0 D}
  U^\dff_D\fp{\sgI}{\ell(\om)}
  \\
  U^\dff_D(x)
  &= U^\dff_N(x)
  =2\int_0^\infty\d u \f{u^3 e^{-u^2}}{u^4+x^4}
  \end{split}
\end{align}
where the asymptotic behavior for $x\to 0$ and $x\to\infty$ behavior is the same as for the noise function (equations (\ref{eq:dff:noise_app}-\ref{eq:dff:noise_app_2})).
\\

\noindent
{\bf 3D Result}
In three dimensions we find dissipation kernel:
\[
  \dissfn(\om)
  =
  \f{\kbt}{4\pi^3 \rh_0D\sgI}
  \int_0^\infty\d u \f{u^4e^{-u^2}}{u^4+(\sgI/\ell)^4}
\]
which we write in terms of a scaling portion and a universal integral function, which is the same as the noise function for 3D diffusion:
\begin{align}
  \label{eq:dfff:diss_raw}
  \tn{(3D)}
  &&
  \begin{split}
  \dissfn(\om)
  &=
  \f{\kbt}{8\pi^{5/2} \rh_0D\sgI}
  U^\dfff_D\fp{\sgI}{\ell(\om)}
  \\
  U^\dfff_D(x)
  &= U^\dfff_N(x)
  =
  \f{2}{\sqrt{\pi}}\int_0^\infty\d u \f{u^4e^{-u^2}}{u^4+x^4}
  \end{split}
\end{align}
where just like the 3D noise universal function, we have that $U^\dfff_N(x\ll 1) = 1$ is a small correction that can be ignored when the sender size is smaller than the characteristic lengthscale.

\subsection{Energy Cost per Bit}
Here we combine our results for the diffusive signal (eq. (\ref{eq:dff:signal_raw}-\ref{eq:dfff:signal_raw})), the noise (eq. (\ref{eq:dff:noise_raw}-\ref{eq:dfff:noise_raw})), and the dissipation (eq. (\ref{eq:dff:diss_raw}-\ref{eq:dfff:diss_raw})) into our expression for the energetic cost per bit (eq. (\ref{eq:cost_per_bit})).
\\

\noindent
{\bf 2D Result}
We find that the energetic cost of diffusively propagating information in 2 dimensions at frequency $\om$ in $\kbt$/bit is given by:
\begin{align}
  \label{eq:dff:cost_raw}
  (2D)
  &&
  \costfn^\dff =
    \log 2
    \f{U^\dff_N\fp{\sgO}{\ell(\om)}U^\dff_D\fp{\sgI}{\ell(\om)}}
      {U^\dff_S\fp{r}{\ell(\om)}}
\end{align}
We note that diffusion in 2D is quite different from all the other studied systems.
This energetic cost has no scaling form;
without peering into the universal functions $U^\dff_S$, $U^\dff_N$, $U^\dff_D$,
we learn nothing about how the energetic cost scales with transmission distance and sender/receiver sizes.
Additionally, its universal functions do not go to $1$ at small lengthscales.
Therefore, we need to look at their asymptotic behavior to get a full picture.

We found earlier that the asymptotic behavior of the signal, noise and dissipation correction functions were given by equations (\ref{eq:dff:signal_app}-\ref{eq:dff:signal_app_2}) and (\ref{eq:dff:noise_app}) with $U^\dff_D=U^\dff_N$.
Below the characteristic lengthscale, in the regime $\sg\ll r \ll \ell$, this yields:
\begin{align}
  \begin{split}
    \costfn^\dff
      &\sim\log 2
      \f{\l(\ga + \ln\fp{\sgI^2}{\ell^2}\r)\l(\ga + \ln\fp{\sgO^2}{\ell^2}\r)}
        {\l(\ln\fp{r}{\ell}\r)^2 + \fp{\pi}{4}^2}
      \\
      &\approx
      4\log 2\f{\ln\fp{\sgI}{\ell}\ln\fp{\sgO}{\ell}}
         {\l(\ln\fp{r}{\ell}\r)^2}
  \end{split}
\end{align}
Above the characteristic lengthscale, in the regime $\sg\ll \ell\ll r$, we find
\begin{align}
  \begin{split}
    \costfn^\dff
      &\sim\log 2
      \f{\l(\ga + \ln\fp{\sgI^2}{\ell^2}\r)\l(\ga + \ln\fp{\sgO^2}{\ell^2}\r)}
        {\f{\pi\ell}{2r}e^{-\sqrt{2}r/\ell}}
      \\
      &\approx
      \f{8\log 2}{\pi}
      \ln\fp{\sgI}{\ell}\ln\fp{\sgO}{\ell}\f{r}{\ell}e^{\sqrt{2}r/\ell}
  \end{split}
\end{align}
Thus the energetic cost per bit of sending a signal behaves like $\ln(\sgI)\ln(\sgO)/\ln(r)^2$ when the transmission distance is below the characteristic lengthscale.
As usual, above the characteristic lengthscale, the energetic cost becomes exponential in the transmission distance.
\\

\noindent
{\bf 3D Result}
Combining the expressions we found for the signal (eq. (\ref{eq:dfff:signal_raw})), noise (eq. (\ref{eq:dfff:noise_raw})), and dissipation (eq. (\ref{eq:dfff:diss_raw})) associated with diffusive signaling in 3D, we find the energetic cost in $\kbt$/bit of signaling is given by:
\begin{align}
  \label{eq:dfff:cost_raw}
  (3D)
  &&
  \costfn^\dfff =
    \f{4\log 2}{\pi}
    \f{r^2}{\sgI\sgO}
    \f{U^\dfff_N\fp{\sgO}{\ell(\om)}U^\dfff_D\fp{\sgI}{\ell(\om)}}
      {U^\dfff_S\fp{r}{\ell(\om)}}
\end{align}
We may then use the asymptotic form we found for the noise/dissipation function (equation (\ref{eq:dfff:noise_scaling})) in the regime $\sg\ll\ell$ to obtain the scaling form of the energetic cost per bit to signal using diffusion in 3D:
\begin{align}
  \label{eq:dfff:cost_scaling}
  \tn{for $\sg\ll \ell$}
  &&
  \costfn^\dfff =
    \f{4\log 2}{\pi}
    \f{r^2}{\sgI\sgO}
    e^{\sqrt{2}r/\ell}
\end{align}
Where we may ignore the contribution $e^{\sqrt{2}r/\ell}$ when the transmission distance is much smaller than the characteristic lengthscale.

\subsection{Connection to Modification-Diffusive Communication}
\label{app:sec:diffusion:modify-diffuse}
Here we briefly talk about the connection between the injection-diffusive system described above where the sender injects/removes signal molecules and modification-diffusive systems where the sender modifies the state of carrier molecules to communicate the signal.
A canonical example of the latter system is the CheY/CheY-P signaling pathway in E coli, where the sender, the kinase CheA, converts the protein CheY to CheY-P through the hydrolysis of ATP.
CheY-P then diffuses through the cytoplasm to the cellular motors, thus communicating the activity state of CheA.

In the injection-diffusive system discussed in this text, the dynamics in the presence of the input signal are given by: $\p_t\lm = D\div^2\lm + J$.
We denote fields related to modified and unmodified carrier molecules using subscripts $a$ and $b$.
Then the dynamics of modification-diffusive systems can be written as:
\begin{align*}
  \p_t\lm_a
  &= D\div^2\lm_a + (\rh-\rh_a)J_a - (\rh - \rh_b)J_b
  \\
  &= D\div^2\lm_a + (\<\rh_b\> - \lm_a)J_a - (\<\rh_a\> - \lm_b)J_b
\end{align*}
where here, $\rh$ describes the total density of modified+unmodified particles, $\rh_a$/$\rh_b$ describe the density of modified/unmodified particles, $J_a$ describes the signal's activation process, and $J_b$ describes the signal's deactivation process.

In the weak-signal regime, the non-linear coupling $(\<\rh_b\>-\lm_a)J_a(x,t)$ can be linearized to $\<\rh_b\> J_a$ and likewise for the off-process.
If we then assume that there the process is symmetric in $a$ and $b$ and that there is no wasted energy in the modification-demodification cycle, we recover the original result found for injection-diffusion.
A more general treatment of modification-diffusive signaling is ongoing work the authors are currently pursuing.

\section{Acoustic Signaling}
\label{app:sec:acoustic}

Here we do an explicit analysis of the energetic cost per bit to send information via acoustic waves in 3D propagating either through water or air.
The dynamics are the same in both water and air, however, the damping strength $\tau$ and the wave velocity $c$ is medium dependent.

\subsection{Setup}

A physical wave may be characterized by the vector field $\bb u(\bb x,t)$ describing its 3D displacement in time and space.
We can describe the low-frequency dynamics of compressive acoustic waves using the fractional Kelvin-Voigt equation, which is a standard model of acoustic waves with viscous relaxation~\cite{wismer_finite_2006,holm_unifying_2010,holm_waves_2019}:
\begin{equation}
  \label{eq:ac2:wave-eq-lossy}
  \f{\p^2}{\p t^2}\bb u - c^2\div^2\bb u - c^2\tau^\et\f{\p^\et}{\p t^\et}\div^2\bb u = 0
\end{equation}
where the timescale $\tau$ controls the strength of viscous relaxation, $\et\leq 1$ is a phenomenological fractional power, and $\p^\et/\p t^\et$ is a fractional derivative, which for our purposes is defined by its Fourier transform:
\[
  \mathcal{F}\l\{\f{\p^\et}{\p t^\et} f(t)\r\}(\om) = (i\om)^\et \mathcal{F}\l\{f\r\}(\om)
\]
When $\tau = 0$, eq (\ref{eq:ac2:wave-eq-lossy}) is the regular undamped wave equation.
In distilled water or ideal monatomic gases, $\et=1$, which is the classical result obtained by Stokes.
In impure media such as blood or seawater, we find that $\et\in\{0.2,1.0\}$~\cite{holm_waves_2019}.
Throughout all of the following analysis, we consider $\et$ in that range.
To first order, the local change in density and pressure are linearly related and given by:
\begin{equation}
  \begin{split}
  \lm &\equiv \rh(x,t) - \rh_0 = - \rh_0\div\cdot\bb u
  \\
  p' &\equiv p(x,t) - p_0 = c^2\lm
  \end{split}
\end{equation}
We also note that we can define the local field velocity $\bb v$.
It obeys the usual continuity equation, but not momentum conservation due to the viscous damping:
\begin{align*}
  \bb v \equiv \f{\p \bb u}{\p t}
  &&
  \p_t\lm = -\rh_0\div\cdot\bb v
  &&
  \p_t(\rh_0\bb v)\neq -\div p
\end{align*}
We choose $\lm$, the density displacement, to play the role of the signal carrying medium to make the analysis of this system symmetric with the other signaling systems studied here.
However, we could just as easily choose $\lm$ to be the pressure fluctuations, $p'$, and we would obtain the exact same results because pressure fluctuations and density fluctuations are equivalent up to a factor of $c^2$.
We will eventually take the divergence of both sides of eq. (\ref{eq:ac2:wave-eq-lossy}) to obtain the dynamics in terms of $\lm$.
However, for now we keep it fully general.

The input and output are defined in terms of $\lm$ in much the same way as with diffusive and electrical signaling.
The sender, $I$, located at the origin injects a particle current $I(t)$ spread out over a Gaussian sphere of radius $\sgI$:
\begin{align}
  J(\bb x,t) = P(\bb x)I(t) = \f{e^{-\bb x^2/2\sgI^2}}{(2\pi\sgI^2)^{3/2}} I(t)
  &&
  \tn{(kg/s/m$^3$)}
\end{align}
The receiver, $O$, located at some position $\bb r$ measures the excess particles in a Gaussian sphere of radius $\sgO$:
\begin{align}
  O(t) = \int\d^3\bb x\, e^{-(\bb x-\bb r)^2/2\sgO^2}\lm(\bb x,t)
  &&
  \tn{(kg)}
\end{align}

\subsection{Parameters}

Electrical signaling was characterized by the RC-velocity $\al/c$.
Diffusive signaling was characterized by the diffusion constant $D$.
This system is more complicated because it has two degrees of freedom, the wave velocity $c$ and the damping parameter $\tau$, which controls the strength of the attenuation.

For signals sent at frequency $\om$, there are two fundamental parameters: the (radian) wavelength $\ell_c$ and the complex dimensionless damping strength $\nu$
\begin{align}
  \label{eq:ac2:parameters-1}
  \ell_c \equiv \f{c}{\om}
  &&
  \nu \equiv i^\et(\tau\om)^\et
\end{align}
The lengthscale $\ell_c$ is just the wavelength divided by $2\pi$.
The dimensionless parameter $\nu$ controls how quickly a signal wave attenuates relative to its period.
Its complex form comes from considering the Fourier transform of the damping term:
\[
  \FF\l\{\tau^\et\f{\p^\et}{\p t^\et} f\r\}(\om)
  = (i\tau\om)^\et \FF\l\{f\r\}(\om)
  = \nu \FF\l\{f\r\}(\om)
\]
We write $\nu$ in terms of its real and imaginary components:
\begin{align}
  \label{eq:ac2:parameters-nu}
  \begin{split}
  \nu &\equiv (i\tau\om)^\et = \nu' + i\nu''
  \\
  \nu' = \cos\fp{\et\pi}{2}(\tau&\om)^\et
  \qquad
  \nu'' = \sin\fp{\et\pi}{2}(\tau\om)^\et
  \end{split}
\end{align}
where $\nu''$ will be particularly important.
From our analysis we will find that there are two characteristic lengthscales which can be expressed as combinations of $\ell_c$ and $\nu''$:
\begin{align}
  \label{eq:ac2:parameters-char}
  \ell_\sg \equiv \ell_c\nu''
  &&
  \ell_r \equiv \f{\ell_c}{\nu''}
\end{align}
The lengthscale $\ell_\sg$ is a lengthscale that will pop out of our analysis (equations (\ref{eq:ac2:diss-function-scaling}-\ref{eq:ac2:diss-app-2})).
The scaling form that the energetic cost of signaling takes depends on the relative value of $\sgI$ and $\ell_\sg$.
The lengthscale $\ell_r$ is the characteristic attenuation lengthscale, analogous to the characteristic lengthscale $\ell$ from the other systems studied.
We later show that the signal attenuates like (see eq. (\ref{eq:ac2:signal-approx})):
\[
  |\ch^{OI}| \propto e^{-r/2\ell_r}
\]
Thus, at transmission distances above $\ell_r$, the energetic cost begins to rise exponentially.
We note that when $\et=1$, this reduces to $\ell_r = \f{c}{\om^2\tau}$, which is the classical formula for the inverse attenuation coefficient, which holds for distilled water and ideal monatomic gases~\cite{holm_waves_2019,lin_bulk_2017}.
This lengthscale $\ell_r$ is more commonly referred to as the inverse acoustic attenuation coefficient $\al$~\cite{holmes_temperature_2011,lin_bulk_2017,dukhin_bulk_2009,bass_atmospheric_1990}, which is typically defined via the attenuation function $e^{-r\al}$.
Thus we can express the relationship between the usual attenuation coefficient $\al$ and $\ell_r$ as:
\begin{equation}
  \label{eq:ac2:acoustic-attenuation-relation}
  \al
  = \f{1}{2\ell_r}
  = \f{\tau^\et\sin(\et\pi/2)}{2c}\om^{1+\et}
  = \al_0\om^{1+\et}
\end{equation}
In section \ref{app:sec:acoustic:parameter-values}, we use this relationship to extract the relevant values for the parameters of this system from empirical measurements of $\al$.

In analyzing the scaling form of the energetic cost of sending information (the cost in the efficient regime) (equations (\ref{eq:ac2:cost-per-bit-raw}-\ref{eq:ac2:cost-per-bit-scaling})), we will work in the regime:
\begin{align*}
  |\nu| \ll 1
  &&
  \sgI,\sgO \ll \ell_c
  &&
  \sgI \ll \ell_\sg
  &&
  0.2 \leq \et \leq 1
\end{align*}
This regime will be justified in section \ref{app:sec:acoustic:parameter-values}.
However, as usual, to generate the plots shown in the main text, we use the exact solution.

\subsection{Parameter Values}
\label{app:sec:acoustic:parameter-values}
The motivation for this section is driven by results we obtain later about the noise and dissipation of the system.
But since it's related to the previous section, we include it here.

The scaling behavior of the noise and dissipation of this system end up depending on the size of the sender $\sgI$ and receiver $\sgO$ relative to the lengthscales:
\begin{align*}
  \sgI \overset{?}{\ll} \ell_\sg = \ell_c\nu''
  &&
  \sgO \overset{?}{\ll} \ell_c(\nu'')^{1/3}
\end{align*}
where $\nu''=\sin\fp{\et\pi}{2}(\tau\om)^\et$ is the imaginary part of the damping parameter.
These lengthscales come from equations (\ref{eq:ac2:noise-function-scaling}-\ref{eq:ac2:noise-app-2}) and (\ref{eq:ac2:diss-function-scaling}-\ref{eq:ac2:diss-app-2}).
Here we show that the regime relevant to intracellular signaling is the regime where the sender and receiver sizes are smaller than these lengthscales.
Since $\nu\ll 1$ in this analysis, this means that $\ell_\sg \ll \ell_c(\nu'')^{1/3}$.
As we will eventually assert that $\sgI\ll \ell_\sg$, $\sgO\ll \ell_c(\nu'')^{1/3}$, this means we can restrict our consideration to $\ell_\sg$, since it is the smaller of the two lengthscales.
It also turns out that the scaling form of the energetic cost of acoustic signaling in this regime depends on $\ell_\sg^2$ (eq. (\ref{eq:ac2:cost-per-bit-scaling})), so it's important to estimate its value.

We can obtain $\ell_\sg$ by first finding $\tau^\et$ by looking at empirical data for the acoustic attenuation coefficient which obeys eq. (\ref{eq:ac2:acoustic-attenuation-relation})
\begin{align*}
  \al
  = \f{1}{2\ell_r}
  = \f{\tau^\et\sin(\et\pi/2)}{2c}\om^{1+\et}
  = \al_0 \om^{1+\et}
\end{align*}
In experiments on blood, fits produce $\al_0=0.14$ $\tn{(dB (MHz)$^{-(1+\et)}$(cm)$^{-1}$)}$, $\et=0.21$~\cite{holm_waves_2019}.
Thus, we can extract $\tau^\et$
\begin{widetext}
\begin{align*}
  \tau^\et\sin\fp{\et\pi}{2} = 2c\al_0
  \approx
    2
    \l(1.5\cdot 10^{5}\,\tn{(cm/s)}\r)
    \l(\f{0.14}{8.686\,(2\pi\cdot 10^6)^{1+\et}}\, \fp{\tn{Np}}{\tn{(rad/s)$^{1+\et}$ (cm)}}\r)
  \approx
  2.9\cdot 10^{-5}\, s^\et
\end{align*}
\end{widetext}
In the paper, we consider frequencies $f\in\{10^{-3}, 10^4\}$ Hz.
This puts the range of  $\ell_\sg$ at:
\[
  \ell_\sg = c\om^{\et-1}\tau^\et\sin\fp{\et\pi}{2}\in \{7\,\tn{$\mu$m}, 2\,\tn{m}\}
\]
Thus, for signaling in intracellular environments, even at the highest frequencies considered, the relevant limit is:
\begin{align*}
  \sgI \ll \ell_\sg \leq 7\,\tn{$\mu$m}
  &&
  \sgO \ll \ell_\sg \ll \ell_\sg(\nu'')^{-2/3}
\end{align*}
As a sanity check, we also find that $\nu$ for the considered frequencies has the range:
\[
  |\nu| = (\om\tau)^\et \in\{ 3\cdot 10^{-5}, 9\cdot 10^{-4} \} \ll 1
\]
So we indeed find that $|\nu|\ll 1$.

Ultimately the goal of including acoustic signaling in this paper is to show that it is an energetically inefficient way to communicate information in intracellular environments.
As far as the authors are aware, no measure of acoustic attenuation in cytoplasm has been published and so values were used for blood.
Blood was chosen because it's a saline biological fluid with a weaker attenuation than other biological media, and thus sets a {\it weak} lower bound.
If one were to produce the plots using values from other tissues, acoustic signaling would be even more inefficient.

\subsection{Dynamics}
Much as with the other systems, we need to compute $\ch^{\lm h}$, the susceptibility of the wave density, and $\ch^{\lm I}$, the response of the wave density to the input signals.
We may then use the fluctuation dissipation theorem to compute the equilibrium fluctuations in $O(t)$ from $\ch^{\lm h}$~\cite{tong_lectures_2012}.

To compute the susceptibility of $\lm$ we need to add an artificial external field $h$ coupled to $\lm$ in the energy density.
This requires us to write down the energy of an acoustic wave, add the term $-h(x,t)\lm(x,t)$, and then observe how this modifies the dynamics given by eq. (\ref{eq:ac2:wave-eq-lossy}).
The Lagrangian density for sound energy in 3D is given by (see eq. 6.17 from~\cite{morse_theoretical_1968}):
\begin{align*}
  \LL = \f{1}{2}\rh(x,t)\bb v(x,t)^2 - \f{1}{2\rh_0c^2}p(x,t)^2
  + h(x,t)\lm(x,t)
\end{align*}
where $h$ is the artificial field described above.
Expanding about equilibrium and keeping only terms to 2nd order in the fluctuation fields:
\begin{align*}
  \LL \approx
    &\f{1}{2}\rh_0\bb v(\bb x,t)^2
    -
    \f{p_0^2}{2\rh_0c^2}
    -
    \f{p_0}{\rh_0c^2}p'(\bb x,t)
    \\
    &-
    \f{1}{2\rh_0c^2}p'(\bb x,t)^2
    + h(x,t)\lm(x,t)
\end{align*}
We can ignore the second term since it's constant and the third term since it integrates to zero. We now rewrite the Lagrangian in terms of the displacement field $\bb u$:
\begin{align}
  \label{eq:ac2:lagrangian}
  \LL = \f{\rh_0}{2}\l[
    \fp{\p \bb u}{\p t}^2
    -
    c^2(\div\cdot\bb u)^2
    -
    2h(x,t)(\div\cdot\bb u)
  \r]
\end{align}
We may ignore the $\rh_0$ since it won't affect the Euler-Lagrange equations, which produce:
\begin{gather*}
  \f{\p\LL}{\p u_x}
  =
  \p_t\fp{\p\LL}{\p(\p_t u_x)}
  +
  \bb\div\cdot\fp{\p\LL}{\p(\bb\div u_x)}
  \\
  0
  =
  \p^2_t u_x
  -
  \p_x\l( c^2\div\cdot\bb u + h(x,t)\r)
\end{gather*}
This gives us the lossless wave-equation with an additional dependence on the field $h$:
\begin{align*}
  \p_t^2\bb {u} = c^2\div^2\bb u + \div h(x,t)
\end{align*}
Adding back the viscous damping term from eq. (\ref{eq:ac2:wave-eq-lossy}), we have the lossy dynamics in the presence of an external field:
\begin{align*}
  \p_t^2\bb u
  =
  c^2\div^2\bb u
  +
  c^2\tau^\et \p_t^\et\div^2\bb u
  +
  \div h(x,t)
\end{align*}
Then by applying $-\rh_0\bb\div\cdot$ to both sides, we obtain the dynamics in terms of $\lm=-\rh_0\div\cdot\bb u$:
\begin{align}
  \label{eq:ac2:dynamics-h-no-signal}
  &&
  \p_t^2\lm
  =
  c^2\div^2\lm
  +
  c^2\tau^\et \p_t^\et\div^2\lm
  -
  \rh_0\div^2 h
  &&
\end{align}
Equation (\ref{eq:ac2:dynamics-h-no-signal}) represents the dynamics in the absence of an input signal.
The input signal $J(\bb x,t)=P(\bb x)I(t)$ modifies the density fluctuations via:
\begin{align*}
  \p_t\lm(\bb x,t) &= \p_t\lm_{\tn{eq}}(\bb x, t) + J(\bb x,t)
  \\
  \p_t^2\lm(\bb x,t) &-  \p_t^2\lm_{\tn{eq}}(\bb x, t) = \p_t J(\bb x,t)
\end{align*}
The linear contribution of this input signal to the dynamics is obtained by simply adding the missing term $\p_t J(\bb x,t)$ to eq. (\ref{eq:ac2:dynamics-h-no-signal}):
\begin{align}
  \label{eq:ac2:dynamics-full}
  \p_t^2\lm
  =
  c^2\div^2\lm
  +
  c^2\tau^\et\p_t^\et\div^2\lm
  -
  \rh_0\div^2 h
  +
  \p_t J
\end{align}
In Fourier space this becomes:
\begin{align*}
  \l[(ck)^2-\om^2 + (i\tau\om)^\et(ck)^2\r]\lm(\bb k,\om)
  =
  \rh_0 k^2 h(\bb k,\om)
  +
  i\om J(\bb k,\om)
\end{align*}
From this we can obtain the response functions $\ch^{\lm h}$ and $\ch^{\lm I}$, defined by $\<\lm(\bb k,\om)\>=\ch^{\lm h}(\bb k,\om)h(\bb k,\om)$ and $\<\lm(\bb k,\om)\>=\ch^{\lm I}(\bb k,\om)I(\om)$:
\begin{equation}
  \begin{split}
  \label{eq:ac2:response-functions}
  \ch^{\lm h}(\bb k,\om) = \f{\rh_0}{c^2} \f{k^2}{k^2-\ell_c^{-2} + \nu k^2}
  \\
  \ch^{\lm I}(\bb k,\om) = \f{i\om}{c^2}\f{e^{-k^2\sgI^2/2}}{k^2-\ell_c^{-2} + \nu k^2}
  \end{split}
\end{equation}
where $\ell_c=c/\om$ is the radian wavelength and $\nu=(i\tau\om)^\et$ is the complex damping parameter (eq. (\ref{eq:ac2:parameters-1})).

\subsection{Signal}
\label{app:sec:acoustic:signal}
Here we compute $|\ch^{OI}(\om)|^2$, the gain of the signaling network.
We start with the response function of $\lm$ to $J$: $\ch^{\lm J}(\bb k,\om) = \f{i\om}{c^2}\f{1}{k^2-(\om/c)^{2} + \nu k^2}$.
Plugging this into eq. (\ref{eq:lemma:output_response_function_3d}) gives:
\begin{align*}
  \ch^{OI}
  =
  \sqrt{\f{2}{\pi}}\f{i\om}{c^2}\f{\sgO^3}{r}
  \int_0^\infty\d u\f{u\sin u}{u^2-(r/\ell_c)^2+\nu u^2}
\end{align*}
This can be analytically solved, allowing us to write the transmission coefficient in terms of a scaling portion and correction function:
\begin{equation}
  \label{eq:ac2:signal-raw}
  \begin{split}
  \l|\ch^{OI}(\om)\r|^2
  &=
  \f{\pi}{2}\fp{\om\sgO^3}{c^2 r}^2
  \l|U^\ac_S\l(\f{r}{\ell_c}, \nu\r)\r|^2
  \\
  U^\ac_S\l(\t z, \nu\r)
  &\equiv
  \f{e^{-i\t z/\sqrt{1+\nu}}}{1+\nu}
  \end{split}
\end{equation}
where we use notation $\t z=r/\ell_c$ for the transmission distance normalized by the radian wavelength.
The function $U^\ac_S$ is a bit trickier than the universal functions for other systems because it is a function of two parameters.
In the regime $|\nu|=(\tau\om)^\et\ll 1$, we can expand the expression in the exponential of $U^\ac_S$ with $|\nu|$ as a small parameter:
\begin{align*}
  e^{\f{-i\t z}{\sqrt{1+\nu}}}
  &=
  \exp\l[-i\t z + \f{1}{2}i\t z\nu + \OO(\nu^2)\r]
  \\
  \l|e^{\f{-i\t z}{\sqrt{1+\nu}}}\r|
  &=
  \exp\l[-\f{1}{2}\t z\nu'' + \OO(\nu^2)\r]
\end{align*}
This gives us the approximate behavior of the correction function:
\begin{equation}
  \label{eq:ac2:attenuation}
  \l|U^\ac_S(\t z,\nu)\r|^2
  \approx
  e^{-\t z\nu''}
  =
  e^{-r/\ell_r}
\end{equation}
with $\ell_r$ being defined from eq. (\ref{eq:ac2:parameters-char}).
This exponential falloff is precisely why we defined $\ell_r$ as we did.
Plugging this into the transmission coefficient expression we obtain the $|\nu|\ll 1$ expression for the gain of the signal:
\begin{align}
  \label{eq:ac2:signal-approx}
  \tn{for $|\nu|\ll 1$}
  &&
  \l|\ch^{OI}(\om)\r|^2
  \approx
  \f{\pi}{2}\fp{\om\sgO^3}{rc^2}^2
  e^{-r/\ell_r}
\end{align}
Thus when the damping parameter $|\nu|=(\om\tau)^{\et}$ is small and the transmission distance is smaller than the characteristic lengthscale ($r\ll \ell_r$), we can ignore the contribution of the correction function $U^\ac_S$.

\subsection{Noise}
\label{app:sec:acoustic:noise}
The power spectrum of the equilibrium fluctuations in the density, $S^\lm(\bb k,\om)$, is given by applying the fluctuation dissipation theorem~\cite{tong_lectures_2012} to $\ch^{\lm h}$:
\begin{align*}
  S^\lm(\bb k,\om)
  &= -\f{2\kbt}{\om}\Im{\ch^{\lm h}(\bb k,\om)}
  \\
  &= -\f{2\kbt\rh_0}{\om c^2}\Im{
    \f{k^2}{k^2-\ell_c^{-2}+\nu k^2}
  }
\end{align*}
By using eq. (\ref{eq:lemma:output_power_spectrum}) we can then show that the equilibrium fluctuations in the output signal are given by:
\begin{align*}
  S^O(\om)
  =
  -\f{\rh_0}{\bt}\f{8\pi\sgO^3}{\om c^2}\Im{
    \int_0^\infty \f{u^4 e^{-u^2}\d u}{u^2-(\sgO/\ell_c)^2+\nu u^2}
  }
\end{align*}
which we write in terms of a scaling portion and an integral correction function:
\begin{align}
  \label{eq:ac2:noise-raw}
  \begin{split}
  S^O(\om) &=
    2\pi^{3/2}
    \f{\rh_0}{\bt}
    \f{\sgO^3\nu''}{c^2\om}
    U^\ac_N\l(\f{\sgO}{\ell_c},\nu\r)
    \\
    U^\ac_N(\t y,\nu) &\equiv -\f{4}{\sqrt{\pi}\nu''}\Im{
      \int_0^\infty\d u\f{u^4 e^{-u^2}}{u^2-\t y^2 + \nu u^2}
    }
  \end{split}
\end{align}
where $\nu''=\sin\fp{\et\pi}{2}\nu$ is the imaginary part of the damping parameter and $\t y=\sgO/\ell_c$ denotes the receiver radius normalized by the radian wavelength.
We have chosen this normalization of $U^\ac_N$ so that we will find $U^\ac_N=1$ in the regime of interest.
The noise correction function $U^\ac_N$ has an analytic solution:
\begin{equation}
  \label{eq:ac2:noise-function-exact}
  \begin{split}
  U^\ac_N(\t y,\nu) = &-\f{1}{\nu''}\tn{Im}\l[
    \f{1}{1+\nu}
    +
    \f{2 \t y^2}{(1+\nu)^2}
  \r.
  \\
  &\l.
    -
    \f{2\sqrt{\pi}i \t y^3}{(1+\nu)^{5/2}}w\l(-\f{\t y}{\sqrt{1+\nu}}\r)
  \r]
  \end{split}
\end{equation}
where $w(z)$ is the Faddeeva function, defined by (see 7.6.3 from~\cite{olver_nist_2010}):
\begin{equation}
  \label{eq:ac2:faddeeva}
  w(z)=e^{-z^2}\erfc(-iz) = \sum_{n=0}^\infty \f{(iz)^n}{\Gamma(n/2 + 1)}
\end{equation}
We can expand $U^\ac_N$ using $|\nu|$, $\t y$ as small parameters:
\begin{equation}
  \label{eq:ac2:noise-function-scaling}
  \begin{split}
  U^\ac(\t y,\nu) \approx
  &\tn{Im}\l[
    i^\et
    \f{\nu}{\nu''}
    +
    2\sqrt{\pi}i\f{y^3}{\nu''}
  \r.
  \\
  &\l.
    +
    \OO(\nu)
    +
    \OO(\t y^2)
    +
    \tn{real}
  \r]
  \\
  \approx
    &1
    +
    2\sqrt{\pi}\f{\t y^3}{\nu''}
  \end{split}
\end{equation}
Thus for $0.2\leq \et\leq 1$, we have two regimes for the noise based on the size of $\t y^3/\nu''$.
When the receiver is small ($\sgO \ll \ell_c(\nu'')^{1/3} \ll \ell_c$), the noise obeys:
\begin{align}
  \label{eq:ac2:noise-app-1}
  U^\ac_N \sim 1
  &&
  S^O(\om)
    &\sim
    2\pi^{3/2}
    \f{\rh_0}{\bt}
    \f{\sgO^3\nu''}{c^2\om}
\end{align}
However, when the receiver is larger ($\ell_c(\nu'')^{1/3} \ll \sgO \ll \ell_c$), the noise obeys:
\begin{align}
  \label{eq:ac2:noise-app-2}
  U^\ac_N \sim \f{2\sqrt{\pi}\sgO^3}{\nu''\ell_c^3}
  &&
  S^O(\om)
    &\sim
    4\pi^{2}
    \f{\rh_0}{\bt}
    \f{\sgO^6\om^2}{c^5}
\end{align}
In the parameter regime discussed in section \ref{app:sec:acoustic:parameter-values}, the relevant scaling behavior is equation (\ref{eq:ac2:noise-app-1}).

\subsection{Dissipation}
\label{app:sec:acoustic:dissipation}
The linearized energy of an acoustic system relative to the zero point is given by (see eq. (\ref{eq:ac2:lagrangian})):
\[
  H = \f{1}{2}\int\d^3\bb x\,\l[
    \rh_0\bb v(\bb x,t)^2
    +
    \f{c^2}{\rh_0}\lm(\bb x,t)^2
  \r]
\]
Thus the dissipation induced by the input signal $J(\bb x,t)$ is given by:
\begin{align*}
  \l\<\diss\r\>
  &= \l\< \f{\p H}{\p\bb v}\fp{\p\bb v}{\p t}_{J}\r\>
    +
    \l\< \f{\p H}{\p\lm}\fp{\p\lm}{\p t}_{J}\r\>
  \\
  &=
  \l\< \f{\p H}{\p\lm}\fp{\p\lm}{\p t}_{J}\r\>
  \\
  &=
  \f{c^2}{\rh_0}
  \l\<\lm(\bb x,t)J(\bb x,t)\r\>
\end{align*}
Then, by using eq. (\ref{eq:lemma:dissipation_kernel}) with $\CC=c^2/\rh_0$, we can show that the dissipation kernel is given by:
\begin{align*}
  \dissfn(\om)
  =
  -\f{\om}{4\pi^3\rh_0\sgI}\Im{
    \int_0^\infty\d u\f{u^2 e^{-u^2}}{u^2-(\sgI/\ell_c)^2 + \nu u^2}
  }
\end{align*}
which we can write in terms of a scaling portion and universal integral function:
\begin{align}
  \label{eq:ac2:diss-raw}
  \begin{split}
  \dissfn(\om)
  &=
  \f{\om\nu''}{8\pi^{5/2}\rh_0\sgI}
  U^\ac_D\l(\f{\sgI}{\ell_c},\nu''\r)
  \\
  U^\ac_D\l(\t x, \nu\r)
  &\equiv
  -
  \f{2}{\sqrt{\pi}\nu''}
  \Im{
    \int_0^\infty\f{\d u\,u^2 e^{-u^2}}{u^2-\t x^2+\nu u^2}
  }
  \end{split}
\end{align}
where $\nu''=|\nu|\sin\fp{\et\pi}{2}$ is the imaginary part of the damping component and
$\t x = \sgI/\ell_c$ denotes the sender radius normalized by the radian wavelength.
We have chosen the normalization of $U^\ac_D$ so that we will find $U^\ac_D=1$ in the relevant regime.
The dissipation correction function $U^\ac_D$ has analytic solution:
\begin{equation}
  \label{eq:ac2:diss-function-exact}
  \begin{split}
  U^\ac_D(\t x,\nu) &= -\f{1}{\nu''}\tn{Im}\biggl[
  \\
  &\l.
    \f{1}{1+\nu}
    -
    i\f{\sqrt{\pi}\t x}{(1+\nu)^{3/2}}w\l(-\f{\t x}{\sqrt{1+\nu}}\r)
  \r]
  \end{split}
\end{equation}
where $w(z)$ is again the Faddeeva function (eq. (\ref{eq:ac2:faddeeva})).
We can expand $U^\ac_D$ using $\nu$, $\t x$ as small parameters (see section \ref{app:sec:acoustic:parameter-values}):
\begin{equation}
  \label{eq:ac2:diss-function-scaling}
  \begin{split}
  U^\ac_D(\t x,\nu)
  &\approx \tn{Im}\biggl[
  \\
    i^\et\f{|\nu|}{\nu''}
    &+ i\sqrt{\pi}\f{\t x}{\nu''}
    +\OO(\nu)+\OO(\t x) + \tn{real}
  \\
  \biggr]
  &\approx 1 + \sqrt{\pi}\f{\t x}{\nu''}
  \end{split}
\end{equation}
Thus for $0.2\leq \et \leq 1$, we have two regimes for the dissipation based on the size of $\t x /\nu''$.
When $\sgI$ is small compared to $\ell_\sg=\ell_c\nu''$ ($\sgI \ll \ell_\sg = \ell_c\nu''\ll \ell_c$), the dissipation obeys:
\begin{align}
  \label{eq:ac2:diss-app-1}
  U^\ac_D \sim 1
  &&
  \dissfn(\om) &\sim \f{\om\nu''}{8\pi^{5/2}\rh_0\sgI}
\end{align}
When $\sgI$ is large compared to $\ell_\sg$ ($\ell_\sg = \ell_c\nu'' \ll \sgI \ll \ell_c$), the dissipation obeys
\begin{align}
  \label{eq:ac2:diss-app-2}
  U^\ac_D \sim \f{\sqrt{\pi}\sgI}{\nu''\ell_c}
  &&
  \dissfn(\om) &\sim \f{\om^2}{8\pi^2\rh_0c}
\end{align}
where, as discussed in section \ref{app:sec:acoustic:parameter-values}, the relevant regime for intracellular signaling is equation (\ref{eq:ac2:diss-app-1}).

\subsection{Energy Cost per Bit}
\label{app:sec:acoustic:cost}
Here we combine our results for the acoustic signal (eq. (\ref{eq:ac2:signal-raw})), the noise (eq. (\ref{eq:ac2:noise-raw})), and the dissipation (eq. (\ref{eq:ac2:diss-raw})) into our expression for the cost (eq. (\ref{eq:cost_per_bit})).
We find that the energetic cost of signaling, in $\kbt$/bit, is given by:
\begin{align}
  \label{eq:ac2:cost-per-bit-raw}
  \costfn^\ac
  &=
    \f{2\log 2}{\pi}
    \f{r^2\ell_\sg^2}{\sgI\sgO^3}
    \f{U^\ac_N\l(\f{\sgO}{\ell_c},\nu\r)U^\ac_D\l(\f{\sgI}{\ell_c},\nu\r)}
      {\l|U^\ac_S\l(\f{r}{\ell_c},\nu\r)\r|^2}
\end{align}
with $\ell_\sg=\nu''\ell_c$, $\ell_c=c/\om$, $\nu''=\Im{(i\tau\om)^\et}$.

\subsubsection{Intracellular Environments}
In section \ref{app:sec:acoustic:parameter-values}, we argued that the regime relevant to intracellular signaling is the regime $\sgI\ll \ell_\sg = \ell_c\nu''$, $\sgO\ll\ell_c(\nu'')^{1/3}$.
In this regime, the noise and dissipation functions can be ignored ($U^\ac_N=U^\ac_D=1$).
Thus the cost in this regime in $\kbt$/bit is given by:
\begin{align}
  \label{eq:ac2:cost-per-bit-scaling}
  \tn{(for $\sgI,\sgO\ll \ell_\sg$)}
  &&
  \costfn^\ac
  &\sim
    \f{2\log 2}{\pi}
    \f{r^2\ell_\sg^2}{\sgI\sgO^3}
    e^{r/\ell_r}
\end{align}
where $\ell_\sg=\ell_c\nu''$ and $\ell_r=\ell_c/\nu''$ come from eq. (\ref{eq:ac2:parameters-char}).

Just as with electrical signaling and diffusive signaling in 3D, we find that there is an exponential dependence of the cost on $r/\ell_r$, the ratio of the transmission distance and the characteristic attenuation lengthscale.
However, unlike these systems, we find that the scaling form of the energetic cost per bit (eq. (\ref{eq:ac2:cost-per-bit-scaling})) depends on the system constants, which enter through the lengthscale parameter:
\[
  \ell_\sg = \ell_c\nu'' = \f{c}{\om}(\tau\om)^\et\sin\fp{\et\pi}{2}
\]
where $0 < \et \leq 1$ is a phenomenological fractional power dependent on the nature of viscous relaxation in the media under consideration.
For distilled water and monatomic gases, $\et=1$ and we obtain $\ell_\sg = c\tau$.
In ocean water, $\et \approx 0.37$~\cite{ainslie_simplified_1998}.
In ultrasound experiments on blood, $\et\approx 0.21$~\cite{duck_physical_1990,holm_waves_2019}.

\subsubsection{Distilled Water}
In distilled water, we find $\et=1$ and $\tau\sim 10^{-12}$ (s).
What happens to the energetic cost of signaling in this regime?
In section \ref{app:sec:acoustic:parameter-values}, we argued that $\sgI\ll \ell_\sg$, $\sgO\ll\ell_c(\nu'')^{1/3}$ in intracellular environments.
In distilled water, this is no longer true.
Since $\et=1$, we have $\nu''=\nu = \tau\om$.
This means that the important lengthscales which dictate the scaling form of the noise and dissipation take on the values:
\begin{align*}
  \ell_\sg = \ell_c\nu'' = c\tau \approx 1.5\,\tn{nm}
  &&
  \ell_c(\nu'')^{1/3} \in\{94\,\tn{$\mu$m}, 4\,\tn{m}\}
\end{align*}
This means that while we still have $\sgO\ll \ell_c(\nu'')^{1/3}$ for sub-cellular receivers, but we no longer should expect that $\sgI\ll \ell_\sg$.
This means we can obtain the energetic cost of acoustic signaling in this regime by
replacing the dissipation equation (\ref{eq:ac2:diss-app-1}) with (\ref{eq:ac2:diss-app-2}) in the energetic cost.
Doing this yields:
\begin{align}
  \label{eq:ac2:cost-per-bit-scaling-2}
  \begin{split}
  \tn{(for $\sgO\ll \ell_c(\nu)^{1/3}$, $\ell_\sg\ll \sgI$)}
  \\
  \costfn^\ac_{\tn{distilled}}
  \sim
    \f{2\log 2}{\sqrt{\pi}}
    \f{r^2}{\sgO^2}
    \f{\ell_\sg}{\sgO}
    e^{r/\ell_r}
  \end{split}
\end{align}
In particular, note how this cost compares to the energetic cost of electrical signaling (eq. \ref{eq:es:cost_scaling}) when $\sgI=\sgO$:
\begin{align*}
  \f{\costfn^\ac_{\tn{distilled}}}{\costfn^\el} = \f{2}{\pi^{3/2}}\f{\ell_\sg}{\sgO}
\end{align*}
This means that acoustic signaling will be more efficient in {\it pure} water whenever $\sgO > \ell_\sg\approx 1.5$ nm.
One may ask then why cells bother with constructing electrical transmission lines in neurons when acoustic transmission lines could hypothetically perform much better.
The answer lies in entropy.
It's far easier to produce an electrochemical gradient by pumping charged ions across the membrane than it is to remove all ions from the aqueous environment.
The entropic cost of purifying water is very large at the cellular scale.

\section{Lemmas}
\label{app:sec:lemmas}

\subsection{Output response from medium response}
\statement[Lemma: $\ch^{OI}$ formula]{
  Consider a system in $d$ dimensions with a density field $\lm$ with mean zero, an output signal defined as the density integrated over a Gaussian shell around point $\bb r$, and a time-dependent input signal that is smeared over a Gaussian shell around the origin.
  \begin{align*}
    O(t) &\equiv \int\d^d\bb x\,e^{-(\bb x-\bb r)^2/2\sgO^2}\lm(\bb x,t)
    \\
    J(\bb x, t) &\equiv \f{e^{-\bb x^2/2\sgI^2}}{(2\pi\sgI^2)^{d/2}}I(t)
  \end{align*}
  Suppose also that we have linearized response function $\ch^{\lm J}$ of $\lm$ to $I(t)$ of the form:
  \begin{equation*}
    \<\lm(\bb k,\om)\>
    \equiv
    e^{-k^2\sgI^2/2}
    \ch^{\lm J}(\bb k,\om)
    I(\om)
  \end{equation*}
  Then the response function of $O(t)$ with respect to $I(t)$ is given by:
  \begin{equation}
    \label{eq:lemma:output_response_function}
    \ch^{OI}(\om)
    \approx
    \f{\sgO^d}{(2\pi r^2)^{\f{d}{2}}}
    \int\d^d\bb u\,
      e^{i\bb u\cdot\bb{\hat{r}}}
      \ch^{\lm J}\l(\f{\bb u}{r},\om\r)
  \end{equation}
  where the approximation holds when $\sgI$, $\sgO\ll r$.
  When $\ch^{\lm J}$ has rotational symmetry, the result for $d=2$ becomes:
  \begin{align}
    \label{eq:lemma:output_response_function_2d}
    \ch^{OI}(\om)
    \approx
    \f{\sgO^2}{r^2}
    \int_0^\infty\d u\,
    u J_0(u)\ch^{\lm J}\l(\f{u}{r},\om\r)
  \end{align}
  where $J_0$ is a Bessel function.
  The result for $d=3$ becomes:
  \begin{align}
    \label{eq:lemma:output_response_function_3d}
    \ch^{OI}(\om)
    \approx
    \sqrt{\f{2}{\pi}}
    \f{\sgO^3}{r^3}
    \int_0^\infty\d u\,
    u \sin u\,\ch^{\lm J}\l(\f{u}{r},\om\r)
  \end{align}
}
\noindent
{\bf Proof:}
We start by writing down $O(\om)$ and expanding $\lm(\bb x,\om)$ into $k$ modes:
\begin{align*}
  \<O(\om)\>
    &= \int\d^d\bb x\, e^{-(\bb x-\bb r)^2/2\sgO^2}\<\lm(\bb x,\om)\>
    \\
    &= \int\f{\d^d\bb x\d^d\bb k}{(2\pi)^d}
      \f{e^{i\bb k\cdot\bb x}}{e^{(\bb x-\bb r)^2/2\sgO^2}}
      \<\lm(\bb k,\om)\>
\end{align*}
Solving the $\bb x$ integral yields
\begin{align*}
  \<O(\om)\>
    &= \f{(2\pi\sgO^2)^{d/2}}{(2\pi)^d}
       \int\d^d\bb k
      \f{e^{i\bb k\cdot\bb r}}{e^{k^2\sgO^2/2}}
      \<\lm(\bb k,\om)\>
\end{align*}
Then we insert the linear response formula for $\<\lm(\bb k,\om)\>$:
\begin{align*}
  \<O(\om)\>
    &= \f{(2\pi\sgO^2)^{d/2}}{(2\pi)^d}
       \int\d^d\bb k
      \f{e^{i\bb k\cdot\bb r}}{e^{k^2(\sgI^2+\sgO^2)/2}}
      \ch^{\lm J}(\bb k,\om)I(\om)
\end{align*}
We then change to the dimensionless integration variable $\bb u=|\bb r|\bb k$ and find:
\begin{align*}
  \<O(\om)\>
    &= \f{\sgO^d}{(2\pi r^2)^{d/2}}
       \int\d^d\bb u
      \f{e^{i\bb u\cdot\bb{\hat{r}}}}{e^{\f{u^2}{2}\l(\f{\sgI^2}{r^2}+\f{\sgO^2}{r^2}\r)}}
      \ch^{\lm J}\l(\f{\bb u}{r},\om\r)I(\om)
\end{align*}
where $r=|\bb r|$.
The linear response function $\ch^{OI}$ of $O(\om)$ to $I(\om)$, defined by $\<O(\om)\>=\ch^{OI}(\om)I(\om)$, is everything in the above expression except $I(\om)$.

Whenever $\sgI$, $\sgO$ $\ll r$, we can drop the exponential term $e^{u^2\sg^2/2r^2}$.
This is a very weak assumption:
since $\sgI$ and $\sgO$ are the sender and receiver sizes, the task of sending information is not well defined when $\sgI$, $\sgO$ $\geq r$.
From this approximation we obtain our desired result:
\begin{align*}
  \ch^{OI}(\om)
    &= \f{\sgO^d}{(2\pi r^2)^{d/2}}
       \int\d^d\bb u\,
      e^{i\bb u\cdot\bb{\hat{r}}}
      \ch^{\lm J}\l(\f{\bb u}{r},\om\r)
\end{align*}
In $d=2$ dimensions, when $\ch^{\lm J}(\bb k,\om)$ is rotationally invariant, we can convert to polar coordinates and perform the $\th$ integral:
\begin{align*}
  \ch^{OI}(\om)
  &= \f{\sgO^2}{2\pi r^2}
      \int\d u\, u\ch^{\lm J}\l(\f{u}{r},\om\r)
      \int_0^{2\pi}\d\th\,e^{i u\cos\th}
  \\
  &=
  \f{\sgO^2}{r^2}
  \int_0^\infty\d u\,
  u J_0(u)
  \ch^{\lm J}\l(\f{u}{r},\om\r)
\end{align*}
where $J_0(u)$ is a Bessel function.
Likewise in $d=3$ dimensions (with rotational invariance), the result is obtained by moving to spherical coordinates:
\begin{align*}
  \ch^{OI}(\om)
  &=
  \f{2\pi\sgO^3}{(2\pi r^2)^{3/2}}
  \int_0^\infty\d u\,u^2\ch^{\lm J}\l(\f{u}{r},\om\r)
  \int_0^\pi\d\th\, \sin\th e^{iu\cos\th}
  \\
  &=
  \sqrt{\f{2}{\pi}}\f{\sgO^3}{r^3}
  \int_0^\infty\d u\,
  u\sin u\,
  \ch^{\lm J}\l(\f{u}{r},\om\r)
\end{align*}

\subsection{Output fluctuations from density fluctuations}
\statement[Lemma: $S^O(\om)$ formula]{
  Consider a system in $d$ dimensions with a density field $\lm$ with mean zero and an output signal $O(t)$ defined as the density integrated over a Gaussian shell around a point $\bb r$:
  \begin{align*}
    O(t) \equiv \int\d^d\bb x\, e^{-(\bb x-\bb r)/2\sgO^2}\lm(\bb x, t)
  \end{align*}
  Then the power spectrum of $O(t)$ is related to the power spectrum of $\lm(\bb x,t)$ by:
  \begin{equation}
    \label{eq:lemma:output_power_spectrum}
    S^O(\om)
    =
    \sgO^{d}\int\d^d\bb u\,
    e^{-u^2}
    S^\lm\l(\f{\bb u}{\sgO},\om\r)
  \end{equation}
  When $S^\lm$ has rotational symmetry, the result for $d=2$ trivially becomes:
  \begin{equation}
    \label{eq:lemma:output_power_spectrum_2d}
    S^O(\om)
    =
    2\pi\sgO^{2}
    \int_0^\infty\d u\,
    u e^{-u^2}
    S^\lm\l(\f{u}{\sgO},\om\r)
  \end{equation}
  For $d=3$ it becomes
  \begin{equation}
    \label{eq:lemma:output_power_spectrum_3d}
    S^O(\om)
    =
    4\pi\sgO^{3}
    \int_0^\infty\d u\,
    u^2 e^{-u^2}
    S^\lm\l(\f{u}{\sgO},\om\r)
  \end{equation}
}
\noindent
{\bf Proof:}
We start with step 2 from the proof of lemma (\ref{eq:lemma:output_response_function}):
\begin{align*}
  O(\om)
    &= \f{\sgO^d}{(2\pi)^{d/2}}
       \int\d^d\bb k
      \f{e^{i\bb k\cdot\bb r}}{e^{k^2\sgO^2/2}}
      \lm(\bb k,\om)
\end{align*}
Inserting this into the expectation value $\<O(\om)O(\om')\>$, we find:
\begin{align*}
  \<O(\om)O(\om')\>
    &= \f{\sgO^{2d}}{(2\pi)^d}
       \int\d^d\bb k\d^d\bb k'
      \f{e^{i(\bb k+\bb k')\cdot\bb r}}
        {e^{\f{\sgO^2}{2}(k^2+k'^2)}}
        \<\lm(\bb k,\om)\lm(\bb k',\om')\>
\end{align*}
Next we insert the identity $\<f(\om)f(\om')\>=2\pi\dl(\om+\om')S^f(\om)$ (see eq. 2.54 in~\cite{bialek_biophysics:_2012}) to both sides of the above expression
(where we apply it to both $\bb k$ and $\om$ on the right hand side):
\begin{align*}
  S^O(\om)
    &= \sgO^{2d}
       \int\d^d\bb k\d^d\bb k'
      \f{e^{i(\bb k+\bb k')\cdot\bb r}}
        {e^{\f{\sgO^2}{2}(k^2+k'^2)}}
        \dl^d(\bb k+\bb k')
        S^\lm(\bb k,\om)
    \\
    &= \sgO^{2d}
       \int\d^d\bb k\,
       e^{-k^2\sgO^2}
       S^{\lm}(\bb k,\om)
\end{align*}
We then change to the dimensionless integration variable $\bb u=\sgO\bb k$, and find the desired result:
\begin{align*}
  S^O(\om)
    &= \sgO^d
       \int\d^d\bb u\,
       e^{-u^2}
       S^{\lm}\l(\f{\bb u}{\sgO},\om\r)
\end{align*}

\subsection{Dissipation kernel from response function}
\statement[Lemma: $\dissfn(\om)$ formula]{
  This gives a formula for the dissipation kernel $\dissfn(\om)$, which is essentially a generalization of the real part of electrical impedance for arbitrary systems.
  Consider a system in $d$ dimensions with a density field $\lm$ with mean zero and an input signal $J(\bb x,t)$ defined via a Gaussian smeared density current:
  \[
    J(\bb x,t) = \f{e^{-\bb x^2/2\sgI^2}}{(2\pi\sgI^2)^{d/2}}I(t)
  \]
  Suppose that we have linearized response function $\ch^{\lm J}$ of $\lm$ to $J$, defined by $\<\lm(\bb k,\om)\>\equiv \ch^{\lm J}(\bb k,\om)J(\bb k,\om)$, which is even in $\bb k$ and Hermitian in $\om$:
  \begin{align*}
    \ch^{\lm J}(-u,\om) = \ch^{\lm J}(u,\om)
    &&
    \ch^{\lm J}(u,-\om) = \overline{\ch^{\lm J}}(u,\om)
  \end{align*}
  And finally, suppose that we can write the dissipative work required to produce this input signal in terms of some factor $\CC$, the input current, and the density field:
  \[
    \<\diss(t)\>
    =
    \CC
    \int\d^d\bb x\, \<J(\bb x,t)\lm(\bb x,t)\>
  \]
  Then the dissipation kernel $\dissfn(\om)$ of $S^{I}(\om)$ is given by
  \begin{equation}
    \label{eq:lemma:dissipation_kernel}
    \begin{split}
    \<\diss\> &= \int\d\om\,\dissfn(\om)S^{I}(\om)
    \\
    \dissfn(\om) &=
      \f{\CC}{\sgI^d}\int
      \f{\d^d\bb u\, e^{-u^2}}{(2\pi)^{d+1}}\,
      \Re{\ch^{\lm J}\l(\f{\bb u}{\sgI},\om\r)}
    \end{split}
  \end{equation}
}
\noindent
{\bf Proof:}
We start by expanding in $\bb k$ modes and using Parseval's theorem:
\begin{align*}
  \<\diss\>
  &=
  \CC\int\d^d\bb x\,\<J(\bb x,t)\lm(\bb x,t)\>
  =
  \CC\int\f{\d^d\bb k}{(2\pi)^d}\<J(\bb k,t)\lm(-\bb k,t)\>
  \\
  &=
  \CC\int\f{\d^d\bb k\d\om\d\om'}{(2\pi)^{d}(2\pi)^2}
  e^{i(\om+\om')t}
  \<J(\bb k,\om)\lm(-\bb k,\om')\>
\end{align*}
We now focus on the Fourier space term, which we evaluate by inserting the linear response function:
\begin{align*}
  \<J(\bb k,\om)\lm(-\bb k,\om')\>
  &=
  \<J(\bb k,\om)J(-\bb k,\om')\>
  \ch^{\lm J}(-\bb k,\om')
  \\
  &=
  \<I(\om)I(\om')\>
  e^{-k^2\sgI^2}
  \ch^{\lm J}(\bb k,\om')
\end{align*}
where we have used that $\ch$ is even in $\bb k$ and the Fourier transform of the Gaussian $J(\bb k,\om)=e^{-\bb k^2\sgI^2/2}I(\om)$.
Next we use the identity $\<I(\om)I(\om')\>=2\pi\dl(\om+\om')S^I(\om)$ (see eq. 2.54 in~\cite{bialek_biophysics:_2012}):
\begin{align*}
  \<J(\bb k,\om)\lm(-\bb k,\om')\>
  &=
  2\pi\dl(\om+\om')S^I(\om)
  e^{-k^2\sgI^2}
  \ch^{\lm J}(\bb k,-\om)
\end{align*}
Plugging this into our original expression, we find:
\begin{align*}
  \<\diss\>
  &=
  \CC\int\d\om\,S^I(\om)
  \int\f{\d^d\bb k}{(2\pi)^{d+1}}
  e^{-k^2\sgI^2}
  \ch^{\lm J}(\bb k,-\om)
\end{align*}
To make the integral dimensionless we change variables to $\bb u = \sgI\bb k$,
\begin{align*}
  \<\diss\>
  &=
  \f{\CC}{\sgI^d}
  \int\d\om\,S^I(\om)
  \int\f{\d^d\bb u}{(2\pi)^{d+1}}
  e^{-u^2}
  \ch^{\lm J}\l(\f{\bb u}{\sgI},-\om\r)
\end{align*}
Finally, we make the observation that if $I(t)$ is real, one can check that $S^I(\om)$ is even in $\om$.
From this and the fact that $\ch^{\lm J}$ is Hermitian in $\om$, we can see that only the real portion of $\ch^{\lm J}$ will contribute to the dissipation:
\begin{align*}
  \int_{-\infty}^\infty\d\om\,S^I(\om)\ch^{\lm J}(-\om)
  &=
  \int_{-\infty}^\infty\d\om\,S^I(\om)\Re{\ch^{\lm J}(\om)}
\end{align*}
This allows us to write down the dissipation kernel $\dissfn(\om)$, which is analogous to the real part of electrical impedance:
\begin{align*}
  \dissfn(\om)
  &=
  \f{\CC}{\sgI^d}
  \int\f{\d^d\bb u}{(2\pi)^{d+1}}
  e^{-u^2}
  \Re{\ch^{\lm J}\l(\f{\bb u}{\sgI},\om\r)}
\end{align*}

\end{document}